\documentclass[12pt,a4]{article}
 \usepackage{amssymb,amsfonts,amsmath,color}

 \usepackage[final]{graphicx}
 \usepackage{wrapfig}
 \usepackage[T2A]{fontenc}
 \usepackage[cp1251]{inputenc}
\usepackage{bm}
 \usepackage{epstopdf}
 
 \newcommand{\beq}{\begin{equation}}
 \newcommand{\eeq}{\end{equation}}
 \newcommand{\bi}{\begin{itemize}}
 \newcommand{\ei}{\end{itemize}}
 \newcommand{\bea}{\begin{eqnarray}}
 \newcommand{\eea}{\end{eqnarray}}
 \newcommand{\ban}{\begin{eqnarray*}}
 \newcommand{\ean}{\end{eqnarray*}}
 \newcommand{\barr}{\begin{array}}
 \newcommand{\earr}{\end{array}}

\renewcommand{\theequation}{\arabic{section}.\arabic{equation}}

\begin{document}

\title{Master equation for open quantum systems: Zwanzig-Nakajima projection technique and
the intrinsic bath dynamics}
\author{V.V.~Ignatyuk$^{a)}$, V.G.~Morozov${^b)}$\\
$^{a)}$\textit{Institute for Condensed Matter Physics, Lviv, Ukraine}\\
$^{b)}$\textit{MIREA-Russian Technological University, Moscow, Russia}}

\date{\today}

 \setcounter{section}{0}

\maketitle

\begin{abstract}
	\noindent
The non-Markovian master equation for open quantum systems is obtained by gene\-ra\-li\-za\-tion of the ordinary Zwanzig-Nakajima (ZN) projection technique. To this end, a coupled chain of equations for the reduced density matrices of the bath $\varrho_{B}(t)$ and the system $\varrho_{S}(t)$ are written down. A formal solution of the equation for $\varrho_{B}(t)$, having been inserted in the equation for the reduced density matrix of the system, in the 2-nd approximation in interaction yields a very specific extra term in the generalized master equation. This term, being nonlinear in $\varrho_{S}(t)$, is related to the intrinsic bath dynamics and vanishes in the Markovian limit. To verify the consistence and robustness of our approach, we applied the generalized ZN projection scheme to a simple dephasing model. It is shown that consideration of the lowest order in interaction is insufficient to describe time evolution of the system coherence
adequately. We explain this fact by analyzing the exact and approximate forms of $\varrho_{B}(t)$ and give some hints how to take the dynamic correlations (which originate from the spin-bath coupling) into account.

\end{abstract}
\section{Introduction}

When studying dynamical processes such as a relaxation, decoherence, buildup of correlations due to the interaction of the open quantum system with its environment, at the certain stage of investigation one inevitably faces the questions: do these phenomena exhibit the Markovian behaviour or not?
What is a role of the dynamical correlations in tending of the system to its local equilibrium?

Many physical systems are believed to be described within Markovian approximation, since the coupling to the environment is weak (Born approximation), and/or the correlations in the bath decay quickly with
respect to the typical time scale of the system’s dynamics. However, one can give some examples when memory effects in the bath cannot be neglected, and the Markov assumption is not applicable anymore. This can be due to strong system-environment couplings \cite{4in-FB,5in-FB}, correlations and entanglement in the initial state \cite{6in-FB,7in-FB}, at the heat transport in nanostructures \cite{13in-FB} or because of specific character of the finite reservoirs \cite{9in-FB}.
The last cases are of particular interest since environment of the open quantum system due to its compactness frequently cannot be regarded as a thermal bath \cite{8in-FB}. In such a case, the dynamics of the reservoir $B$ has to be treated (at least, at the initial stage of evolution) on equal footing with that of the $S$-subsystem.

A powerful tool for dealing with such systems is provided
by the projection operator techniques \cite{14in-FB,15in-FB}, which have
been introduced by Nakajima \cite{16in-FB}, Zwanzig \cite{17in-FB}, and Mori
\cite{18in-FB}. This approach has manifested its efficiency at the construction of the generalized master equation and investigation of the non-Markovian dynamics in the initially correlated open quantum systems \cite{Meier-Tannor-1999}, in the spin star systems \cite{Petruccione-2004} and spin baths \cite{Fischer-Breuer-2007}, for a spin coupled to an environment subject to the external field \cite{China-2019} and many others. 

However, ZN  scheme has also some disadvantages \cite{BP-Book}. Though the generic master equation is usually written down up to the 2-nd order in interaction, time convolution in the kinetic kernels is determined by the \textit{full} evolution operators (including an interaction part $V$ of the total Hamiltonian).
It means that one has to take into account the entire series in the coupling constant to provide a regular analysis of the system dynamics. This is a cornerstone of the time-convolutionless equation (TCL) technique \cite{BP-Book}, when one moves from the retarded dynamics to the equations local in time with the time-dependent generators, which include both upward and backward evolution. 
Such method yields a systematic perturbative expansion scheme for the stochastic dynamics
of the reduced system which is valid in an intermediate coupling
regime, where non-Markovian effects are important \cite{Vacchini-Breuer-2010,Breuer-Petruccione-1999,Breuer-2016}.

However, none of the above methods consider the \textit{intrinsic dynamics} of the environment. One certainly has to calculate time evolution of the bath operators before averaging, but that of the reduced density matrix of the environment $\varrho_B(t)$ is usually ignored. Though for the infinite thermal baths this simplification is well justified, some questions about its validity could appear, when one deals with the finite size reservoirs \cite{9in-FB} or tries to investigate the impact of running correlations \cite{particles}. In other words, the open system dynamics has definitely to be in accord with the concept \cite{ZMR,Gemmer-Michel-Physica}, claiming that an evolution towards
local equilibrium is always accompanied by an increase of the system-bath correlations even though the asymptotic values of the observables can strongly differ from those predicted by the Markov-Born approximation \cite{8in-FB}. 

In this paper, we generalize the ordinary ZN projection technique on our way to derive the non-Markovian master equation for the open quantum systems. Initially in Section 2, we start from the coupled set of equations for the reduced density matrices of the system $\varrho_S(t)$ and of the environment $\varrho_B(t)$, respectively. Then in Section 3 we insert a formal solution of the dynamic equation for $\varrho_B(t)$ in the equation for the reduced density matrix of the $S$-subsystem, and in the second order in interaction obtain the generalized master equation. 

The above generalization gives rise to the extra term of a very specific structure:
i) it is nonlinear in $\varrho_S(t)$ and ii) vanishes in the Markovian limit. These two points make the situation very similar to that, which one faces when studying the inset of running correlations: the generic kinetic equations are strongly nonlinear, and the correlational contribution to the collision integral tends to zero in the Markovian limit \cite{ZMR,MorRop01}. 

In Section 4, we compare our results with those following from the standard ZN scheme. To verify the consistence of our approach, we apply in Section 5 the generalized ZN projection scheme to a simple dephasing model \cite{PRA2012} and obtain the kinetic equation for the system coherence. However, the results of the numerical solution of the above equation, which are represented in Section 6, lead us to a disencouraging conclusion: the lowest order in interaction is insufficient to describe time evolution of the system coherence adequately. In Section 7, we give our explanation to this fact by analyzing the exact and approximate forms of the $\varrho_{B}(t)$. In the last Section, we make conclusions and give some hints how to take the running correlations, which originate from the spin-bath coupling, into account.

\section{Basic  equations}

\subsection{Definitions}
Suppose that the composed system under consideration ($S+B$) consists of the open quantum system (subsystem $S$) and its surroundings $B$, which usually can be considered as a thermal bath. The total Hamiltonian of the above mentioned system
 \beq
   \label{Hamilt}
 H(t)=H^{}_{S}(t) + H^{}_{B} + V
 \eeq
 consists of the term $H^{}_{S}(t)$ corresponding to the open quantum system $S$ (which in the general case is allowed to depend on time $t$ due to the action of the eventual external fields), the summand $H^{}_{B}$ related to the bath, and the interaction term $V$.
 
The density matrix $\varrho(t)$ of the composed system obeys the quantum Liouville equation (hereafter we put $\hbar=1$).
 \beq
  \label{Eq-rho}
  \frac{\partial\varrho(t)}{\partial t}= -i\left[H(t),\varrho(t)\right].
 \eeq
 
 For further convenience, let us pass to the interaction picture $\widetilde{\varrho}(t)$ for the total density matrix, which is defined as
 \beq
 \label{rho-IP}
 \widetilde{\varrho}(t)= U^{\dagger}(t)\varrho(t)U(t),
 \eeq
 where we have introduced the unitary evolution operators
  \beq
   \label{U-def}
   \begin{array}{c}
  U(t)=U^{}_{S}(t)U^{}_{B}(t),
  \\[10pt]
  \displaystyle
  U^{}_{S}(t)= \exp^{}_{+}\left\{-i\int^{t}_{0} dt'\,H^{}_{S}(t')\right\},
  \qquad
  U^{}_{B}(t)=\exp\left(-itH^{}_{B}\right).
  \end{array}
  \eeq
In Eq.~(\ref{U-def}), the expression $\exp_+\left\{\ldots\right\}$ denotes a time ordered exponent, which turns into a much simpler form $U^{}_{S}(t)=\exp\left(-itH^{}_{S}\right)$ if the open system Hamiltonian $H_S$ does not depend on time. The unitary operator $U(t)$ and its Hermitian conjugate  counterpart $U^{\dagger}(t)$ obey the evolution equations
 \beq
   \label{U-eq}
  \frac{d U(t)}{dt}=
   -\left\{iH^{}_{S}(t)+iH^{}_{B}\right\}U(t),
   \qquad
  \frac{d U^{\dagger}(t)}{dt}=
   U^{\dagger}(t)\left\{iH^{}_{S}(t)+iH^{}_{B}\right\},
 \eeq
where from it is easy to represent the quantum Liouville equation (\ref{Eq-rho}) in the interaction picture,
 \beq
  \label{Eq-rho-IP}
 \frac{\partial\widetilde{\varrho}(t)}{\partial t}=
 -i\left[\widetilde{V}(t),\widetilde{\varrho}(t)\right].
 \eeq
Hereafter, operators in the interaction picture are defined as
  \beq
    \label{A-IP}
  \widetilde{A}(t)= U^{\dagger}(t)A U(t).
  \eeq
It should be mentioned that if $A^{}_{S}$ ($A^{}_{B}$) is an operator acting in the Hilbert space
  of the subsystem $S$ (bath $B$), then the corresponding interaction representation can be written down as follows:
   \beq
    \label{A-IP-SB}
    \widetilde{A}^{}_{S}(t)=U^{\dagger}_{S}(t)A^{}_{S} U^{}_{S}(t),
    \qquad
   \widetilde{A}^{}_{B}(t)=U^{\dagger}_{B}(t)A^{}_{B} U^{}_{B}(t).
   \eeq

The reduced density matrices for the system $S$ and the
environment are introduced in a usual way,
  \beq
    \label{rho-SB}
    \varrho^{}_{S}(t)= \text{Tr}^{}_{B}\varrho(t),
  \quad
 \varrho^{}_{B}(t)= \text{Tr}^{}_{S}\varrho(t),
  \eeq
  by taking a trace over the environment (the system) variables.
The corresponding time averages for operators $A_S$ ($A_B$) can be introduced as follows:
 \beq
   \label{Ave-rho-SB}
   \begin{array}{l}
  \langle A^{}_{S}\rangle^{t}\equiv
   \text{Tr}^{}_{SB}\left\{A^{}_{S}\varrho(t)\right\}=
   \text{Tr}^{}_{S}\left\{A^{}_{S}\varrho^{}_{S}(t)\right\},
   \\[10pt]
  \langle A^{}_{B}\rangle^{t}\equiv
   \text{Tr}^{}_{SB}\left\{A^{}_{B}\varrho(t)\right\}=
   \text{Tr}^{}_{B}\left\{A^{}_{B}\varrho^{}_{B}(t)\right\}.
   \end{array}
 \eeq

The reduced density matrix $\widetilde{\varrho}^{}_{S}(t)$ of the open quantum system is defined as
  \beq
   \label{IP-rho-S}
  \widetilde{\varrho}^{}_{S}(t)=
  U^{\dagger}_{S}(t)\varrho^{}_{S}(t)U^{}_{S}(t).
  \eeq
Analogously, in the interaction picture the reduced density matrix $\widetilde{\varrho}^{}_{B}(t)$ of the environment takes the form
     \beq
   \label{IP-rho-B}
  \widetilde{\varrho}^{}_{B}(t)=
  U^{\dagger}_{B}(t)\varrho^{}_{B}(t)U^{}_{B}(t).
  \eeq
It is easy to verify that
  \beq
  \label{IP-rho-SB}
 \widetilde{\varrho}^{}_{S}(t)=
  \text{Tr}^{}_{B}\widetilde{\varrho}(t),
  \qquad
 \widetilde{\varrho}^{}_{B}(t)=
  \text{Tr}^{}_{S}\widetilde{\varrho}(t).
  \eeq

\subsection{Equations of motion for density matrices}

To derive evolution equations for the operators $\widetilde{\varrho}_S(t)$ and $\widetilde{\varrho}_B(t)$, we apply the following decomposition for the total density matrix:
     \beq
  \varrho(t)= \varrho^{}_{S}(t)\varrho^{}_{B}(t)+ \Delta \varrho(t),
   \label{rho-decomp}
  \eeq
where the correlation term satisfies the relations
   \beq
   \text{Tr}^{}_{S}\,\Delta \varrho(t)=0,
   \quad
 \text{Tr}^{}_{B}\,\Delta \varrho(t)=0.
    \label{Del-rho-prop}
   \eeq
In the interaction picture, Eqs.~(\ref{rho-decomp})-(\ref{Del-rho-prop}) convert into the similar relations,
       \beq
  \widetilde{\varrho}(t)=
    \widetilde{\varrho}^{}_{S}(t) \widetilde{\varrho}^{}_{B}(t)
    + \Delta  \widetilde{\varrho}(t)
   \label{rho-decomp1}
  \eeq
and
   \beq
 \text{Tr}^{}_{B}\,\Delta  \widetilde{\varrho}(t)=0,
 \qquad
   \text{Tr}^{}_{S}\,\Delta  \widetilde{\varrho}(t)=0.
    \label{IP-Del-rho-prop}
   \eeq

Taking the trace $\text{Tr}^{}_{B}$ of both sides of
Eq.~(\ref{Eq-rho-IP}), we get
  \beq
   \label{IP-eq-S-1}
  \frac{\partial\widetilde{\varrho}^{}_{S}(t)}{\partial t}=
 -i\text{Tr}^{}_{B}\left[\widetilde{V}(t),\widetilde{\varrho}(t)\right].
  \eeq
Using the decomposition (\ref{rho-decomp1}), one can rewrite Eq.~(\ref{IP-eq-S-1}) in the following form:
   \beq
   \label{IP-eq-S-2}
  \frac{\partial\widetilde{\varrho}^{}_{S}(t)}{\partial t}=
  -i\left[\widetilde{V}^{}_{S}(t),\widetilde{\varrho}^{}_{S}(t)\right]
 -i\text{Tr}^{}_{B}
 \left[\widetilde{V}(t),\Delta\widetilde{\varrho}(t)\right],
  \eeq
where
  \beq
   \label{H-int-S}
   \widetilde{V}^{}_{S}(t)=\text{Tr}^{}_{B}
   \left\{
  \widetilde{V}(t)\widetilde{\varrho}^{}_{B}(t)
   \right\}.
  \eeq

In a similar way, having taken a trace $\text{Tr}^{}_{S}$ of Eq.~(\ref{Eq-rho-IP}), one obtains
     \beq
   \label{IP-eq-B-2}
  \frac{\partial\widetilde{\varrho}^{}_{B}(t)}{\partial t}=
  -i\left[\widetilde{V}^{}_{B}(t),\widetilde{\varrho}^{}_{B}(t)\right]
 -i\text{Tr}^{}_{S}
 \left[\widetilde{V}(t),\Delta\widetilde{\varrho}(t)\right],
  \eeq
where
  \beq
   \label{H-int-B}
   \widetilde{V}^{}_{B}(t)=\text{Tr}^{}_{S}
   \left\{
  \widetilde{V}(t)\widetilde{\varrho}^{}_{S}(t)
   \right\}.
  \eeq

To derive the equation of motion for correlational part $\Delta\widetilde{\varrho}(t)$ of the total density matrix, 
let us rewrite Eqs.~(\ref{Eq-rho-IP}), (\ref{IP-eq-S-2}), and
(\ref{IP-eq-B-2}) in the form
 \beq
   \label{IP-eq-rho}
   \frac{\partial \widetilde{\varrho}(t)}{\partial t}=
    -i{\mathcal L}(t)\widetilde{\varrho}(t),
 \eeq
 \beq
  \label{IP-eq-S-L}
  \frac{\partial \widetilde{\varrho}^{}_{S}(t)}{\partial t}=
    -i{\mathcal L}^{}_{S}(t)\widetilde{\varrho}^{}_{S}(t)
     - \text{Tr}^{}_{B}\left\{i{\mathcal L}(t)\,
              \Delta \widetilde{\varrho}(t)\right\},
 \eeq
  \beq
  \label{IP-eq-B-L}
  \frac{\partial \widetilde{\varrho}^{}_{B}(t)}{\partial t}=
    -i{\mathcal L}^{}_{B}(t)\widetilde{\varrho}^{}_{B}(t)
     - \text{Tr}^{}_{S}\left\{i{\mathcal L}(t)\,
              \Delta \widetilde{\varrho}(t)\right\}.
 \eeq
Here we have introduced the Liouville operators ${\mathcal L}(t)$, ${\mathcal L}_S(t)$ and ${\mathcal L}_B(t)$ via the corresponding commutators:
  \beq
   \label{L-def}
  {\mathcal L}(t)A=[\widetilde{V}(t),A],
  \qquad
  {\mathcal L}^{}_{S}(t)A=[\widetilde{V}^{}_{S}(t),A],
  \qquad
  {\mathcal L}^{}_{B}(t)A=[\widetilde{V}^{}_{B}(t),A].
  \eeq
Using the decomposition (\ref{rho-decomp1}) and taking into account Eqs.~(\ref{IP-eq-rho})-(\ref{IP-eq-B-L}), one can obtain the evolution of motion for correlational part of the total density matrix,
  \bea
  \label{eq-del-rho}\nonumber
  & &
 \hspace*{-50pt}
   \frac{\partial \,\Delta\widetilde{\varrho}}{\partial t}=
  \frac{\partial \,\widetilde{\varrho}}{\partial t}
  -\widetilde{\varrho}^{}_{S}\frac{\partial \widetilde{\varrho}^{}_{B}}{\partial t}
 -\widetilde{\varrho}^{}_{B}\frac{\partial \widetilde{\varrho}^{}_{S}}{\partial t}
 \\[10pt]
  & &
  \hspace*{-50pt}
  {}=-i{\mathcal L}\widetilde{\varrho}
  -\widetilde{\varrho}^{}_{S}
  \left\{
 -i{\mathcal L}^{}_{B}\widetilde{\varrho}^{}_{B}
     - \text{Tr}^{}_{S}\left\{i{\mathcal L}\,
              \Delta \widetilde{\varrho}\right\}
  \right\}
 -\widetilde{\varrho}^{}_{B}
  \left\{
 -i{\mathcal L}^{}_{S}\widetilde{\varrho}^{}_{S}
     - \text{Tr}^{}_{B}\left\{i{\mathcal L}\,
              \Delta \widetilde{\varrho}\right\}
  \right\}.
  \eea
The above equation can be written down in a more compact form,
  \beq
   \label{IP-eq-Del-rho}
   \left(
   \frac{\partial}{\partial t}
   +{\mathcal Q}(t)i{\mathcal L}(t){\mathcal Q}(t)
   \right) \Delta \widetilde{\varrho}(t)=
   -{\mathcal Q}(t)i{\mathcal L}(t)
   \widetilde{\varrho}^{}_{S}(t)\widetilde{\varrho}^{}_{B}(t),
  \eeq
  using the superoperators
  \beq
   \label{P-Q}
   {\mathcal Q}(t)=1-{\mathcal P}(t),
  \qquad
 {\mathcal P}(t)A=
 \widetilde{\varrho}^{}_{S}(t)\,\text{Tr}^{}_{S}A
 + \widetilde{\varrho}^{}_{B}(t)\,\text{Tr}^{}_{B}A.
  \eeq
It can be shown that, when acting on operators with
 zero trace, $\text{Tr}A=0$, the superoperator ${\mathcal P}(t)$
  satisfies the relation ${\mathcal P}^{2}(t)={\mathcal P}(t)$, i.e., it is a
  projector. Indeed, we have ($t$ is omitted)
  \ban
 {\mathcal P}^{2}A&=&
  \widetilde{\varrho}^{}_{S}\,\text{Tr}^{}_{S}
  \left(
 \widetilde{\varrho}^{}_{S}\,\text{Tr}^{}_{S}A +
 \widetilde{\varrho}^{}_{B}\,\text{Tr}^{}_{B}A
  \right)
 +
  \widetilde{\varrho}^{}_{B}\,\text{Tr}^{}_{B}
  \left(
 \widetilde{\varrho}^{}_{S}\,\text{Tr}^{}_{S}A +
 \widetilde{\varrho}^{}_{B}\,\text{Tr}^{}_{B}A
  \right)
  \\[8pt]
   {}&=&
 \widetilde{\varrho}^{}_{S}\,\text{Tr}^{}_{S}A +
 2\widetilde{\varrho}^{}_{S}\widetilde{\varrho}^{}_{B}\,\text{Tr}A
  +\widetilde{\varrho}^{}_{B}\,\text{Tr}^{}_{B}A=
  {\mathcal P}A.
   \ean

A formal solution of Eq.~(\ref{IP-eq-Del-rho}) is
 \beq
  \label{Del-rho-formal}
 \Delta\widetilde{\varrho}(t)=
  - i\int^{t}_{0}dt'\,
   {\mathcal U}(t,t'){\mathcal Q}(t'){\mathcal L}(t')
   \widetilde{\varrho}^{}_{S}(t')\widetilde{\varrho}^{}_{B}(t'),
 \eeq
where the superoperator ${\mathcal U}(t,t')$ satisfies the equation of motion
 \beq
   \label{MU-eq}
  \frac{\partial {\mathcal U}(t,t')}{\partial t}=
   -i{\mathcal Q}(t){\mathcal L}(t){\mathcal Q}(t){\mathcal U}(t,t'),
   \qquad {\mathcal U}(t',t')=1,
 \eeq
where we have introduced the evolution operator
 \beq
   \label{MU-expr}
  {\mathcal U}(t,t')=\exp^{}_{+}
  \left\{
  -i\int^{t}_{t'}d\tau\,
  {\mathcal Q}(\tau){\mathcal L}(\tau){\mathcal Q}(\tau)
  \right\}.
 \eeq
Substituting expression (\ref{Del-rho-formal}) into
 Eqs.~(\ref{IP-eq-S-L}) and (\ref{IP-eq-B-L}),
  we arrive at the closed system of equations
 for the reduced density matrices,
 \beq
   \label{IP-eq-S}
  \frac{\partial \widetilde{\varrho}^{}_{S}(t)}{\partial t}=
  -i{\mathcal L}^{}_{S}(t)\widetilde{\varrho}^{}_{S}(t)
  -\int^{t}_{0}dt'\,
   \text{Tr}^{}_{B}
   \left\{
   {\mathcal L}(t){\mathcal U}(t,t'){\mathcal Q}(t'){\mathcal L}(t')
   \widetilde{\varrho}^{}_{S}(t')\widetilde{\varrho}^{}_{B}(t')
   \right\},
 \eeq
 \beq
   \label{IP-eq-B}
  \frac{\partial \widetilde{\varrho}^{}_{B}(t)}{\partial t}=
  -i{\mathcal L}^{}_{B}(t)\widetilde{\varrho}^{}_{B}(t)
  -\int^{t}_{0}dt'\,
   \text{Tr}^{}_{S}
   \left\{
   {\mathcal L}(t){\mathcal U}(t,t'){\mathcal Q}(t'){\mathcal L}(t')
   \widetilde{\varrho}^{}_{S}(t')\widetilde{\varrho}^{}_{B}(t')
   \right\}.
 \eeq
 However, Eq.~(\ref{IP-eq-S}) cannot be considered yet as a master equation for the open quantum system since it depends on the reduced density matrix of the environment. A task to solve Eq.~(\ref{IP-eq-B}) with respect to the density matrix of the environment $\widetilde{\varrho}_B(t)$ seems to be unrealistic because of its complexity. 
Thus the only way is to use some approximations for $\widetilde{\varrho}_B(t)$ obtainable from Eq.~(\ref{IP-eq-B}). 
This is a subject of the next Section.

\setcounter{equation}{0}

\section{Master equation for $\widetilde{\varrho}_S(t)$:
 weak coupling approximation}

Let us start this Section with two assumptions:
  \bi
  \item
  \textit{
  	If the subsystem $S$ is small compared to the environment $B$,
  it is reasonable to suppose that, for sufficiently small
   time $t$, the state of the environment is close to
   $\varrho^{}_{B}(0)$.
 Based on the above assumption, we write the solution of
  Eq.~(\ref{IP-eq-B}) as}
   \beq
    \label{IP-rho-B-lin}
  \widetilde{\varrho}_{B}(t)\approx \varrho^{}_{B}(0)
 - \int^{t}_{0}dt'\,i{\mathcal L}^{}_{B}(t')\varrho^{}_{B}(0).
   \eeq
   \item
  \textit{On the right-hand side of Eq.~(\ref{IP-eq-S}), we set ${\mathcal U}(t,t')=1$ and}
   $\widetilde{\varrho}^{}_{B}(t')=\varrho^{}_{B}(0)$.
  \ei
The above assumptions mean that we restrict the equation for $\widetilde{\varrho}_B(t)$ by the first order in interaction. Based on these approximations, we get
 \beq
   \label{IP-eq-S-weak}
   \frac{\partial \widetilde{\varrho}^{}_{S}(t)}{\partial t}=
  -i{\mathcal L}^{}_{S}(t)\widetilde{\varrho}^{}_{S}(t)
  -\int^{t}_{0}dt'\,
   \text{Tr}^{}_{B}
   \left\{
    {\mathcal L}(t){\mathcal Q}^{(0)}(t'){\mathcal L}(t')
   \widetilde{\varrho}^{}_{S}(t')\varrho^{}_{B}(0)
   \right\},
 \eeq
where
  \beq
    \label{Q-lin}
  {\mathcal Q}^{(0)}(t)A= A
  -  \widetilde{\varrho}^{}_{S}(t)\,\text{Tr}^{}_{S}A
  - \varrho^{}_{B}(0)\,\text{Tr}^{}_{B}A.
  \eeq
After some manipulations with the r.h.s. of Eq.~(\ref{IP-eq-S-weak}), which are described in detail in the Appendix A, we obtain the final master equation in the interaction picture: 
 \bea
\label{master-fin}
& &
\hspace*{-40pt}
\frac{\partial \widetilde{\varrho}^{}_{S}(t)}{\partial t}=
-i{\mathcal L}^{(0)}_{S}(t)\widetilde{\varrho}^{}_{S}(t)
-
\int^{t}_{0}dt'\,
\left[ R^{}_{S}(t,t'),
\widetilde{\varrho}_{S}(t)-\widetilde{\varrho}_{S}(t')\right]
\nonumber\\[10pt]
& &
{}- \int^{t}_{0}dt'\,
\left\{
\text{Tr}^{}_{B}
\left\{
{\mathcal L}(t)
{\mathcal L}(t')\widetilde{\varrho}^{}_{S}(t')\varrho^{}_{B}(0)
\right\}
- {\mathcal L}^{(0)}_{S}(t){\mathcal L}^{(0)}_{S}(t')
\widetilde{\varrho}_{S}(t')
\right\}.
\eea

Some comments on the structure of master equation (\ref{master-fin}) are quite pertinent at this stage:
  \bi
  \item
 The first term on the r.h.s. vanishes if $\varrho^{}_{B}(0)$ is
  an \textit{equilibrium\/} density matrix of the bath and
   $$
   \text{Tr}^{}_{B}\left\{V\varrho^{}_{B}(0)\right\}=0.
   $$
   \item
   In the above case the term ${\mathcal L}^{(0)}_{S}(t){\mathcal L}^{(0)}_{S}(t')
     \widetilde{\varrho}_{S}(t')$ also vanishes. It is shown in Section 5 that the terms with ${\mathcal L}^{(0)}_S$ describe an influence of the initial correlations in the open quantum systems. In particular, a non-integral term contributes to the non-dissipative properties, whereas the integral term forms a new channel of dissipations dealt with the initial preparation of the system.
   \item
   The second term on the r.h.s. is quite interesting:

    1) It vanishes in the Markovian limit   
      \footnote{One should distinguish two kinds of approximation for the integrand. The first one looks as
      	\linebreak
    	$
    	\int_0^t \mathcal{K}(t-\tau)f(\tau)d\tau\approx
    	\int_0^{\mathbf t} \mathcal{K}(t-\tau)d\tau f(t).
    	$
    	This approximation is often called \textit{the Markovian approximation of the 1-st kind} \cite{review} or 
    	the \textit{Redfield approximation} (dealt with the corresponding Redfield equation \cite{BP-Book} for the reduced density matrix of the subsystem $S$). 
    	The other approximation,
    	$
    	\int_0^t \mathcal{K}(t-\tau)f(\tau)d\tau \approx
    	\int_0^{\mathbf{\infty}}d\tau \mathcal{K}(t-\tau)d\tau f(t)
    	$
    	is known as \textit{the Markovian approximation of the 2-nd kind} \cite{review}. Very often \cite{ZMR}, it is being positioned just as \textit{the Markovian approximation}. In this paper, we use the Markovian approximation of the 1-st kind only.
    }, when dynamics of the kinetic kernel $R_S(t,t')$ is considered as a fast one in comparison with time evolution of the density matrix; then we set $\widetilde{\varrho}_{S}(t')\approx\widetilde{\varrho}_{S}(t)$.

    2) This term is \textit{nonlinear\/} in
    $\widetilde{\varrho}_{S}$ since the operator $\widetilde{V}^{}_{B}(t')$
     appearing in Eq.~(\ref{R}) depends on
      $\widetilde{\varrho}_{S}(t')$ (see also Eq.~(\ref{H-int-B})).
A nonlinearity appears due to consideration of the intrinsic dynamics of the environment by means of time evolution of the reduced density matrix $\widetilde{\varrho}_B(t)$. There is some resemblance with the basic points following from the quantum kinetic theory \cite{ZMR,MorRop01}, where the running correlations yield nonlinear terms in the kinetic equations, which are vanishing in the Markovian limit. However, it is shown in Section 6 that approximation (\ref{IP-rho-B-lin}) is not sufficient to describe correctly the influence of bath dynamics, and consideration of the high order terms in interactions for $\widetilde{\varrho}_B(t)$ is necessary.
  \ei

 \setcounter{equation}{0}

 \section{Master equation in the Zwanzig-Nakajima scheme}

It would be useful to compare the results of Section 3 with those which follow from the traditional ZN projection scheme. In this scheme one usually uses another decomposition for the total density matrix instead of (\ref{rho-decomp}), 
\beq
\label{Zwanz-decomp-eq}
\varrho(t)= \varrho^{}_{S}(t)\varrho^{}_{B}(0)+ \delta \varrho(t).
\eeq
In fact, decomposition (\ref{Zwanz-decomp-eq}) means that we neglect the intrinsic dynamics of the environment by setting the reduced density matrix of the bath to be equal to its initial value at any time, $\varrho_B(t)\equiv\varrho_B(0)$. It should be noted that in the original ZN scheme \cite{BP-Book} it is supposed that
$[H^{}_{B},\varrho^{}_{B}(0)]=0$, i.e., the bath is initially in
thermal equilibrium. If this initial state differs from thermal
equilibrium, it is natural to write the decomposition
of $\varrho(t)$ in the form
\beq
\label{Zwanz-decomp-gen}
\varrho(t)= \varrho^{}_{S}(t)\varrho^{(0)}_{B}(t)+ \delta \varrho(t),
\eeq
where
\beq
\label{rho-B-0}
\varrho^{(0)}_{B}(t)= U^{}_{B}(t)\varrho^{}_{B}(0)U^{\dagger}_{B}(t)
\equiv
{\rm e}^{-iH^{}_{B}t}\varrho^{}_{B}(0){\rm e}^{iH^{}_{B}t}.
\eeq
Clearly, the decomposition (\ref{Zwanz-decomp-gen}) guarantees that
the correction term $\delta\varrho(t)$ vanishes in the absence of
interaction between the system $S$ and the environment.
In the interaction picture Eq.~(\ref{Zwanz-decomp-gen}) reduces to
         \beq
  \widetilde{\varrho}(t)=
    \widetilde{\varrho}^{}_{S}(t) \varrho^{}_{B}(0)
    + \delta  \widetilde{\varrho}(t).
   \label{Zw-IP-decomp}
  \eeq
It is important to note that (c.f. Eqs.~(\ref{IP-Del-rho-prop}))
     \beq
\label{Zw-IP-Del-rho}
 \text{Tr}^{}_{B}\,\delta  \widetilde{\varrho}(t)=0,
 \qquad
   \text{Tr}^{}_{S}\,\delta  \widetilde{\varrho}(t)\not=0.
   \eeq

Proceeding in a way as in Section 3 (see also Appendix B for details), we obtain
the ZN master equation
 \beq
   \label{Zw-master-I}
   \frac{\partial\widetilde{\varrho}^{}_{S}(t)}{\partial t}=
  - i{\mathcal L}^{(0)}_{S}(t)\widetilde{\varrho}^{}_{S}(t)
 -\int^{t}_{0}dt'\, \text{Tr}^{}_{B}
  \left\{
  {\mathcal L}(t)
  {\mathcal U}^{}_{Z}(t,t'){\mathcal Q}^{}_{Z}\,{\mathcal L}(t')\,
  \widetilde{\varrho}^{}_{S}(t')\varrho^{}_{B}(0)
  \right\}.
 \eeq
 \quad
In the weak coupling limit, this equation can be presented in the form similar to (\ref{master-fin}):
     \bea
     \label{ZW-master-wI}
  & &
  \hspace*{-40pt}
    \frac{\partial \widetilde{\varrho}^{}_{S}(t)}{\partial t}=
  -i{\mathcal L}^{(0)}_{S}(t)\widetilde{\varrho}^{}_{S}(t)
  \nonumber\\[10pt]
   & &
 {}- \int^{t}_{0}dt'\,
  \left\{
  \text{Tr}^{}_{B}
   \left\{
    {\mathcal L}(t)
  {\mathcal L}(t')\widetilde{\varrho}^{}_{S}(t')\varrho^{}_{B}(0)
    \right\}
    - {\mathcal L}^{(0)}_{S}(t){\mathcal L}^{(0)}_{S}(t')
     \widetilde{\varrho}_{S}(t')
    \right\}.
 \eea
 However, in contrast to Eq.~(\ref{master-fin}), the master equation (\ref{ZW-master-wI})
  does not contain the nonlinear term with $R^{}_{S}(t,t')$ due to exclusion of the instrinsic dynamics of the environment from consideration. Moreover, in the Markovian limit,
both master equations are identical. This is quite obvious, since the influence of bath dynamics (considered here in terms of $\widetilde{\varrho}_B(t)$) is expected to manifest itself only at the initial stage of the system evolution.

To check out a consistency and robustness of the generalized projection scheme, presented in Sections 2 and 3, one has to apply the master equation (\ref{master-fin}) to derive the quantum kinetic equations for observables in some simple models (preferably, excactly solvable ones). It is performed in the next Section, where we obtain the dynamic equation for coherence, considering purely dephasing model \cite{PRA2012}. We explore both the generalized master equation (\ref{master-fin}) and its ZN reduction (\ref{ZW-master-wI}) and point out to the evident discrepancy between the data, which follow from the generalized projection scheme, and the exact results \cite{PRA2012} for the dephasing model.

\section{Non-Markovian quantum kinetic equation for	the dephasing model}

\setcounter{equation}{0}

\subsection{The model Hamiltonian}
We consider a simple version of the spin-boson model describing a
two-state system ($S$) coupled to a bath ($B$) of harmonic
oscillators \cite{PRA2012,TMF}. In the spin representation for a
qubit, the total Hamiltonian of the model is written as (in our
units $\hbar=1$)
\bea\label{H}H=H_S+H_B+V=\frac{\omega_0}{2}\sigma_3+\sum\limits_k\omega_k
b^{\dagger}_k b_k+\sigma_3\sum\limits_k(g_k b^{\dagger}_k+g^*_k
b_k),\eea where $\omega_0$ is the energy difference between the
excited state $|1\rangle$ and the ground state $|0\rangle$ of the
qubit, and $\sigma_3$ is one of the Pauli matrices $\sigma_1$,
$\sigma_2$, $\sigma_3$. Bosonic creation and annihilation
operators $b^{\dagger}_k$ and $b_k$ correspond to the $k$-th bath
mode with frequency $\omega_k$.

Since $\sigma_3$ commutes with Hamiltonian (\ref{H}), it does not
evolve in time, $\sigma_3(t)=\sigma_3$. Hence, the interaction
operator can be easily expressed as
\bea\label{tildeV1}\tilde V(t)\equiv\sigma_3\sum\limits_k
{\cal F}_k (t)=\sigma_3\sum\limits_k\left\{g_k(t)
b^{\dagger}_k+g^*_k(t)b_k \right\},
\qquad
g_k(t)=g_k \exp(i\omega_k t).
\eea
It can be also shown, using the identity
$\sigma_3\cdot\sigma_3={\cal I}$, that the interaction operator in the
Heisenberg picture has a similar structure as (\ref{tildeV1}),
 \bea\label{V(t)} V(t)=\exp(i H t)V\exp(-i H
t)=\sigma_3\sum\limits_k {\cal F}_k (t)+C(t),\eea where a
$c$-number term \bea\nonumber
&C(t)=\sum\limits_k\left\{g_k(t)\alpha^*_k(t)+g^*_k(t)\alpha_k(t) \right\}, &
\alpha_k(t)=g_k \frac{1-e^{i\omega_k t}}{\omega_k} \eea should be
omitted at subsequent calculation of commutators.

\subsection{The non-Markovian kinetic equation for coherence}

At the beginning, let us introduce in a usual fashion the spin
inversion operators $\sigma_{\pm}=(\sigma_1\pm i\sigma_2)/2$,
which obey the permutation relations
$\left[\sigma_3,\sigma_{\pm}\right]=\pm 2\sigma_3$. Our task is to
obtain the quantum kinetic equations for the mean values
$\langle\widetilde{\sigma}_{\pm}\rangle_{S}^{t}=\mbox{Tr}_S\{\tilde\rho_S(t)\sigma_{\pm}\}$ dealt with the
system coherence. For simplicity, we consider the case with
$\langle\widetilde{\sigma}_+\rangle_{S}^{t}$, since the equation for $\langle\widetilde{\sigma}_-\rangle_{S}^{t}$ can be easily obtained in a similar manner.

Using the basic equation (\ref{master-fin}) for the reduced density
matrix $\tilde\rho_S(t)$ and taking trace over the system
variables, one can derive the following kinetic equation:
\bea\label{kinet1}\nonumber \frac{\partial
	\langle\widetilde \sigma_{+}\rangle_{S}^{t}}{\partial
	t}=\underbrace{i\left\langle\left[\tilde
	V_S^{(0)}(t),\sigma_+\right]\right\rangle^t_{S}}_{(1)}+\int\limits_0^t
dt'\biggl\{\underbrace{\left\langle\left[R_S(t,t'),\sigma_+\right]\right\rangle^t_{S}}_{(2)}
-\underbrace{\left\langle\left[R_S(t,t'),\sigma_+\right]\right\rangle^{t'}_{S}}_{(2')}\biggr\}\\
-\int\limits_0^t
dt'\biggl\{\underbrace{\left\langle\left\langle\left[\tilde
	V(t),\left[\tilde
	V(t'),\sigma_+\right]\right]\right\rangle_B\right\rangle^{t'}_{S}}_{(3)}+\underbrace{\left\langle\left[\tilde
	V_S^{(0)}(t),\left[\tilde
	V_S^{(0)}(t'),\sigma_+\right]\right]\right\rangle^{t'}_{S}}_{(4)}\biggr\},\eea
where we denote the bath averaging as $\langle\ldots\rangle_{B}\equiv\mbox{Tr}_B\left\{\varrho_{B}(0)\ldots
\right\}$.

Having calculated all commutators in Eq.~(\ref{kinet1}) and performed thermal averaging (see Appendixes C and D for details), we can write down the final kinetic equation for the generalized coherence:
\bea\label{kinetFin}\nonumber
\frac{d\left\langle\widetilde\sigma_+\right\rangle^t_{S}}{dt}=i
A(\psi)\int\limits_0^{\infty}\frac{J(\omega)}{\omega}\cos\omega
t\,dt\left\langle\widetilde\sigma_+\right\rangle^t_{S}\\\nonumber-
i\langle\sigma_3\rangle\int\limits_0^t\Biggl\{\int\limits_0^{\infty}J(\omega)\sin[\omega(t-t')]d\omega\Biggr\}\left(\langle\widetilde\sigma_+\rangle^{t}_{\widetilde
	S}-\langle\widetilde \sigma_+\rangle^{t'}_{S}\right)dt'
\\\nonumber
-\int\limits_0^t\Biggl\{\frac{1}{2}\int\limits_0^{\infty}J(\omega)\coth\left(\frac{\omega}{2
	k_B
	T}\right)\cos[\omega(t-t')]d\omega\\+2(A(\psi)^2+1)\int\limits_0^{\infty}\frac{J(\omega)}{\omega}\cos\omega
t\,d\omega\int\limits_0^{\infty}\frac{J(\omega')}{\omega'}\cos\omega't'\,d\omega'\Biggr\}\left\langle\widetilde\sigma_+
\right\rangle^{t'}_{S}dt'. \eea

 Let us discuss some peculiarities of the quantum
kinetic equation (\ref{kinetFin}). First of all, we remind that
the second term in the r.h.s of (\ref{kinetFin}) vanishes in the Markovian
limit. On the other hand, this term (being imaginary one) along
with the first (``quasi-free'') term should contribute to the qubit
frequency $\omega_0$, yielding the corresponding frequency shift
\cite{PRA2012} \footnote{We can always pass from the mean values in
	the interaction picture to the averages taken with the statistical
	operator $\rho_S(t)$ according the rule
	$\left\langle\sigma^+\right\rangle^t_{\widetilde
		S}\rightarrow\exp(-i\omega_0
	t)\left\langle\sigma^+\right\rangle^t_S$.}.

Secondly, the initial state of the qubit contributes to the quasi-free term as well as to the last summand
in the r.h.s of Eq.~(\ref{kinetFin}) by means of the values $A(\psi)$ and
$\langle\sigma_3\rangle $. 
It is also worthy to emphasize the following point. While the kinetic kernels in the 2-nd and the
3-rd terms have the usual form ${\cal
	K}(t-t')$, this is not true for the last term dealt with the
initial correlations in the ``qubit+bath'' system. It is not surprising: the dynamics of initial correlations does not belong to the class of the stationary processes with a typical convolution dependence of kinetic kernels; the initial correlations are just ageing \cite{Pottier}.

In the third place, the contribution of the initial correlations (see
the last term in the r.h.s. of Eq.~(\ref{kinetFin})) is of the 4-th order in interaction in a complete
correspondence with the results of Ref.~\cite{PRA2012}. Moreover, it is straightforward to show that in the Markovian limit a solution of the kinetic equation (\ref{kinetFin}) completely reproduces the result of Ref.~\cite{PRA2012} for both the qubit energy shift and the total decoherence. 

In the next Section we examine what new occurs with the qubit dynamics if one considers the generic non-Markovian equation (\ref{kinetFin}) for the system coherence.

\section{Numerical solution of the non-Markovian quantum kinetic equation 
%	(\ref{kinetFin})
}  
\setcounter{equation}{0}

In this Section, we present the results of numerical solution of the non-Markovian quantum kinetic equation (\ref{kinetFin}). The qubit-bath coupling is modelled by the spectral function $J(\omega)$ taken in its standard form \cite{PRA2012}
\bea\label{J1(w)}
J(\omega)=\lambda_s\Omega^{1-s}\omega^s \exp(-\omega/\Omega).
\eea
The parameter $\lambda_s\sim |g_k|^2$ in Eq.~(\ref{J1(w)}) denotes a dimensionless coupling constant, while $\Omega$ means the cut-off frequency. Depending on the exponent $s$, we can distinguish three coupling regimes: the sub-Ohmic at $0<s<1$, the Ohmic at $s=1$ and the super-Ohmic at $s>1$. 
We consider the Ohmic and super-Ohmic qubit-bath coupling; besides, we study the qubit dynamics at both low and high temperatures.

\begin{figure*}[htb]
	\centerline{\textbf{\footnotesize{LT}} \hspace{2mm} \includegraphics[height=0.2\textheight,angle=0]{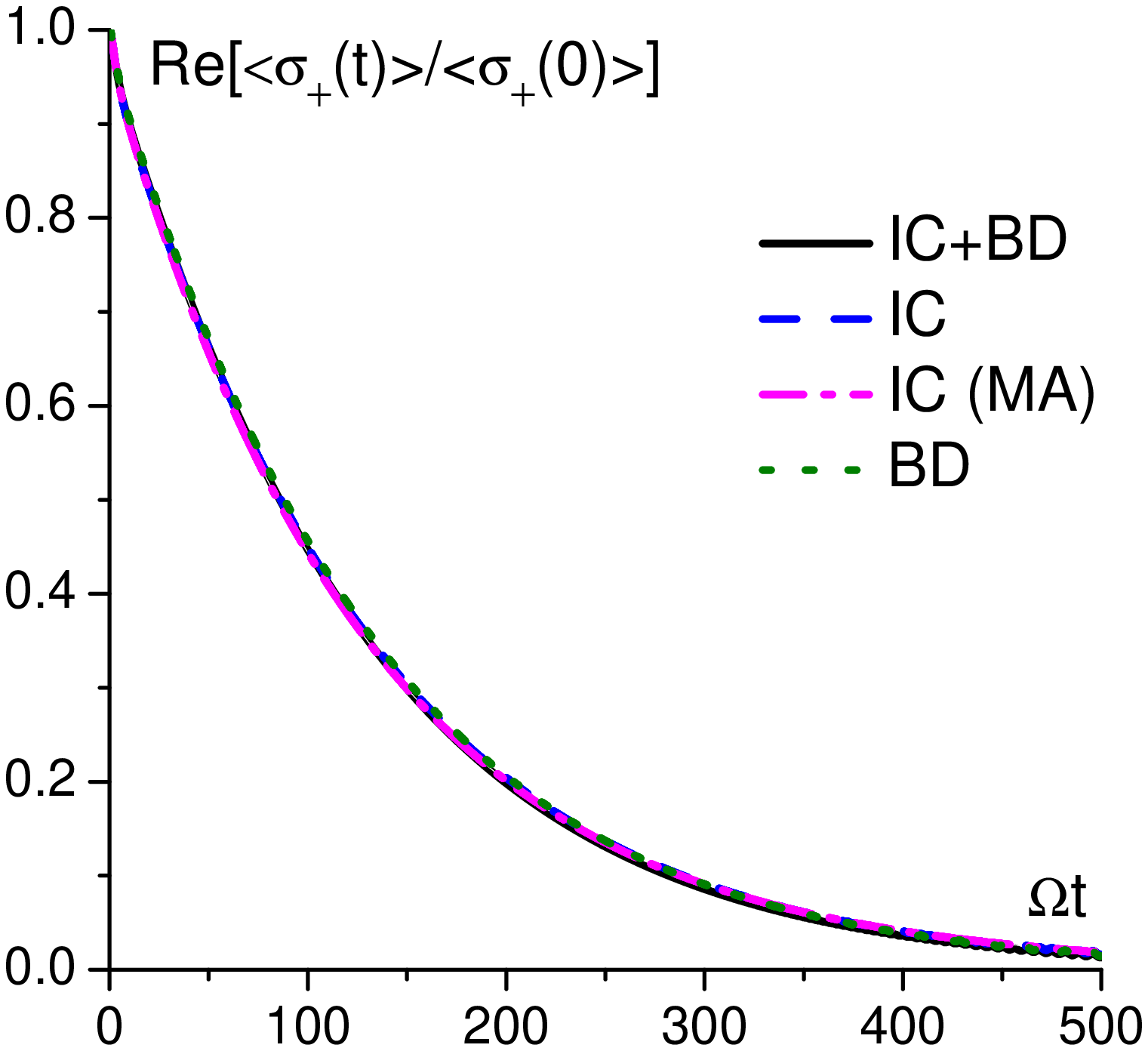}
		\hspace{1cm}
		\includegraphics[height=0.2\textheight,angle=0]{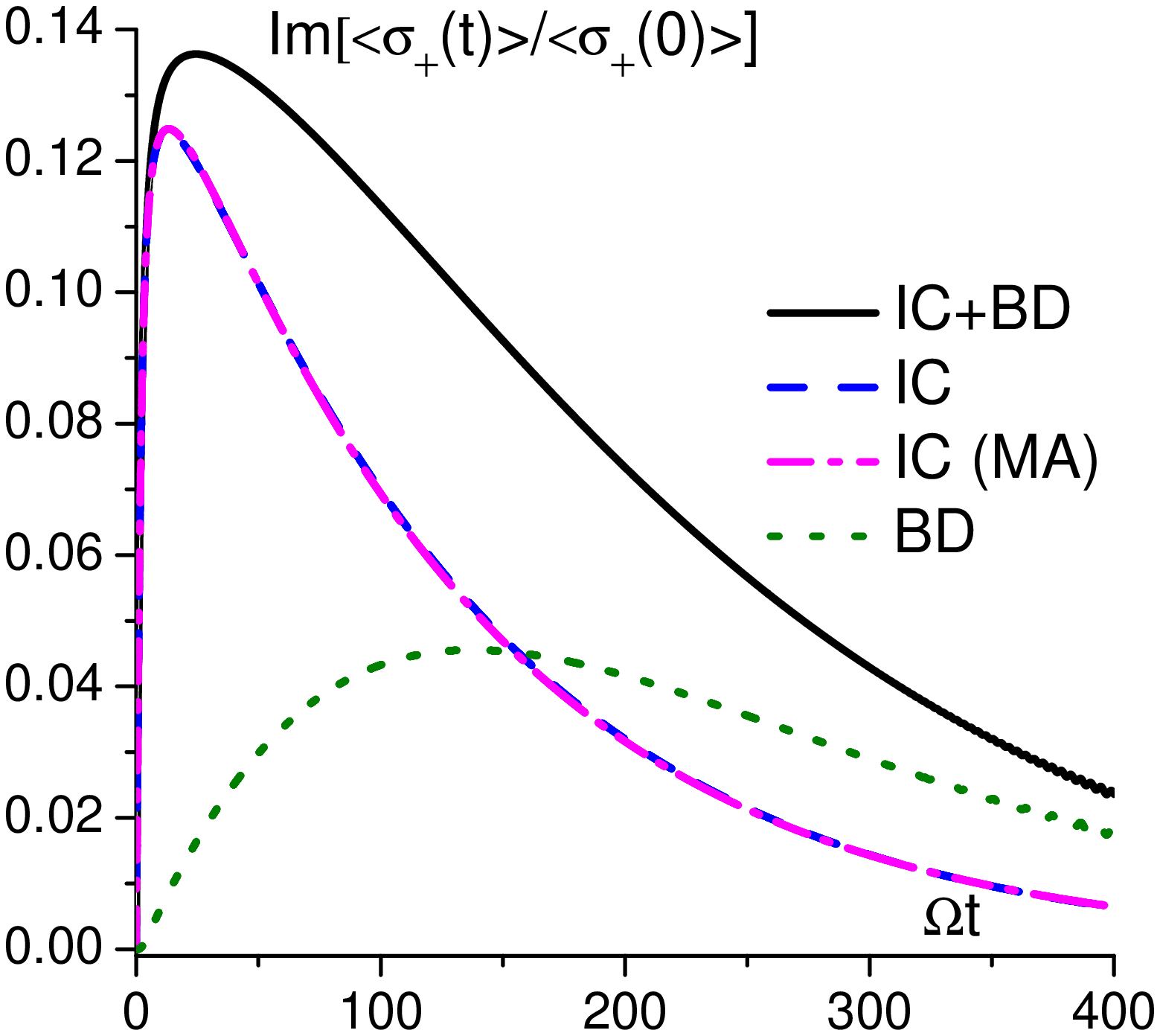}}
	\vspace{0.2cm}
	\centerline{\textbf{\footnotesize{HT}}\hspace{2mm} \includegraphics[height=0.2\textheight,angle=0]{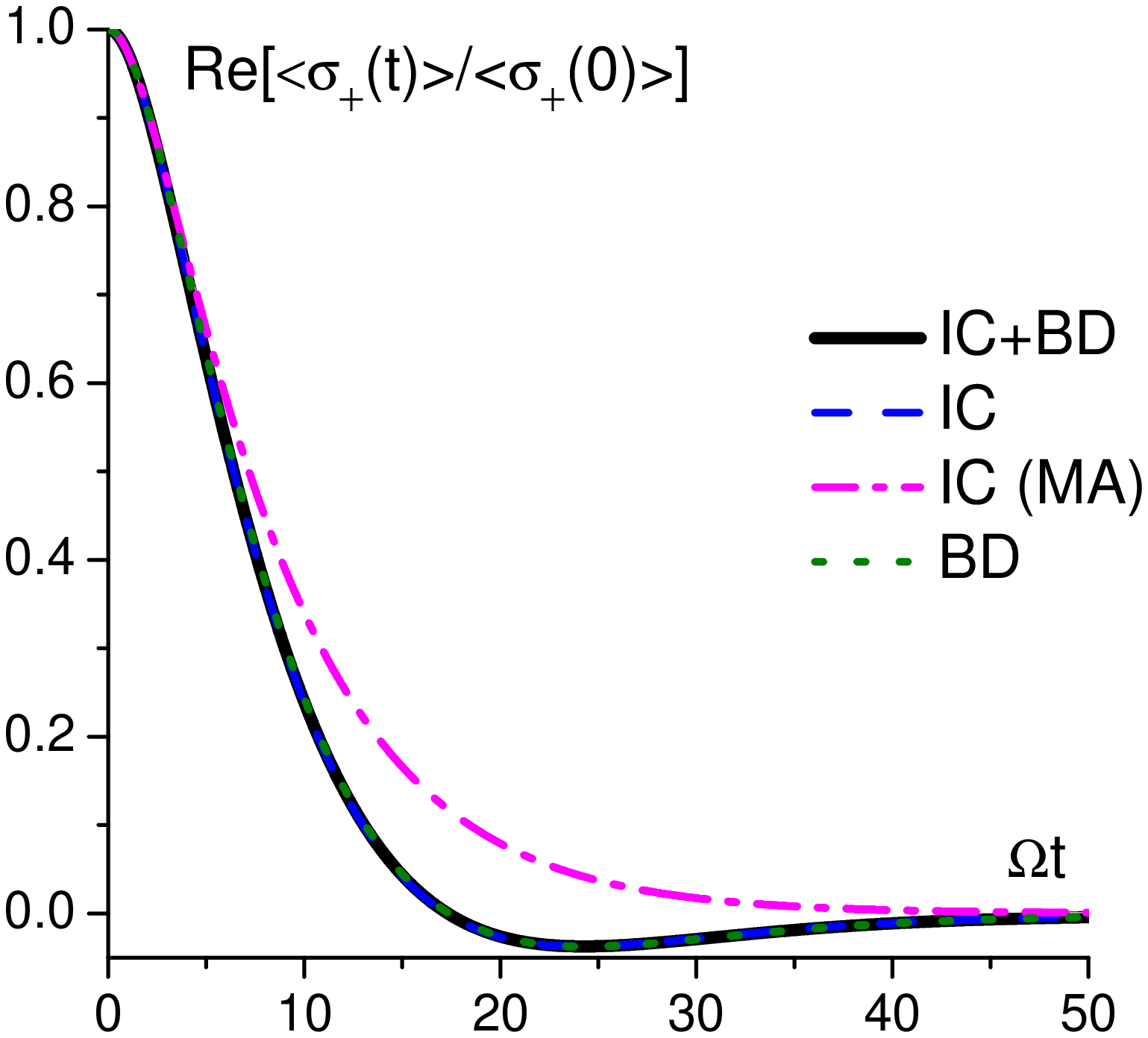}
		\hspace{1cm}
		\includegraphics[height=0.2\textheight,angle=0]{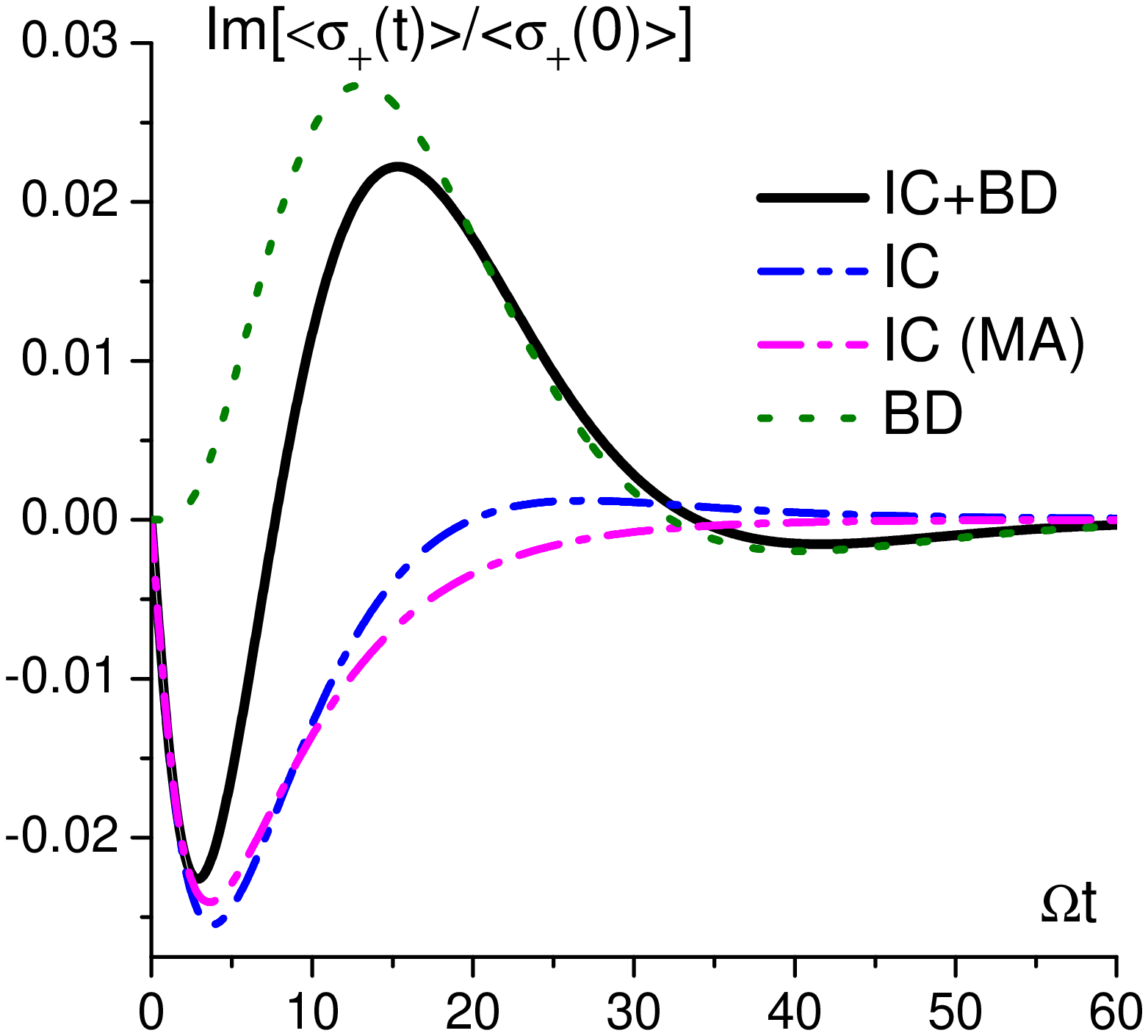}}
	\caption{\small{Time evolution of the real (left column) and imaginary (right column) parts of the normalized coherence $\langle\tilde\sigma_+\rangle^t_S/\langle\tilde\sigma_+\rangle^{t=0}_S$ in the Ohmic coupling regime ($s=1$) at low $T^*=1/\beta\Omega=0.025$ (LT, an upper panel) and high temperature $T^*=1/\beta\Omega=0.5$ (HT, a lower panel). Other system parameters: $\lambda_s=0.1$, $\Omega=1$, $\langle\sigma_3\rangle=1/2$. Solid curves correspond to the solution of the non-Markovian equation (\ref{kinetFin}), when both the initial correlations (IC) and the bath dynamics (BD) are taken into account; dashed (dashed-dotted) curves describe the non-Markovian (Markovian) dynamics when the BD is neglected; dotted curves describe the non-Markovian dynamics when the IC are neglected.}}
\end{figure*}

The results of numerical solution of Eq.~(\ref{kinetFin}) are depicted in Figs.~1 and 2. Here the black solid curves describe the system coherence, when both initial correlations (IC) and bath dynamics (BD), introduced by the reduced density matrix $\widetilde{\varrho}_B(t)$, are taken into account. The blue dashed curves correspond to the case, when only IC are considered (the second term in the r.h.s. of Eq.~(\ref{kinetFin}) is omitted), while the green dotted ones are associated with BD in the absence of IC (the first and the last terms in the r.h.s. of Eq.~(\ref{kinetFin}) are omitted). The magenta dashed-dotted curves are reference ones since they correspond to the Markovian approximation (MA) which, as it has been already said, yields the exact results up to the 2-nd order in interaction \cite{PRA2012}.
It is also obvious that the third term in the r.h.s. of Eq.~(\ref{kinetFin}), which defines the cumulative contribution of the vacuum and thermal fluctuations, has to be always retained either in the non-Markovian approach or in MA.

The Fig.~1 corresponds to the Ohmic coupling regime. It is seen from the upper panel that in the low temperature limit
the real parts of the generalized coherence are very close in all four cases considered. Thus one can conclude that the qubit dynamics is governed mainly by the vacuum and the thermal parts \cite{PRA2012} of the generalized decoherence, while the contribution of the IC and BD is vanishing. The situation becomes somewhat different for the imaginary part of the coherence: taking into account the IC and/or the BD deviates the corresponding curves (solid and dotted) from the reference one (dashed-dotted). It should be also noted that the non-Markovian effects play a minor role in this case since the blue dashed curve (related to the non-Markovian dynamics) almost overlaps with the reference one (dealt with the MA).

In the high temperature limit, the difference between the non-Markovian dynamics and the MA becomes much more pronounced: all curves deviate from the reference one (the magenta dashed-dotted curve) both for the real and the imaginary parts of the generalized coherence. Moreover, consideration of the BD makes this deviation even stronger, as it is clearly seen from the right plot (for the $\mbox{Im}[\langle\sigma_+(t)/\langle\sigma_+(0)\rangle]$).

\begin{figure*}[h]
	\centerline{\textbf{\footnotesize{LT}}\hspace{2mm}\includegraphics[height=0.2\textheight,angle=0]{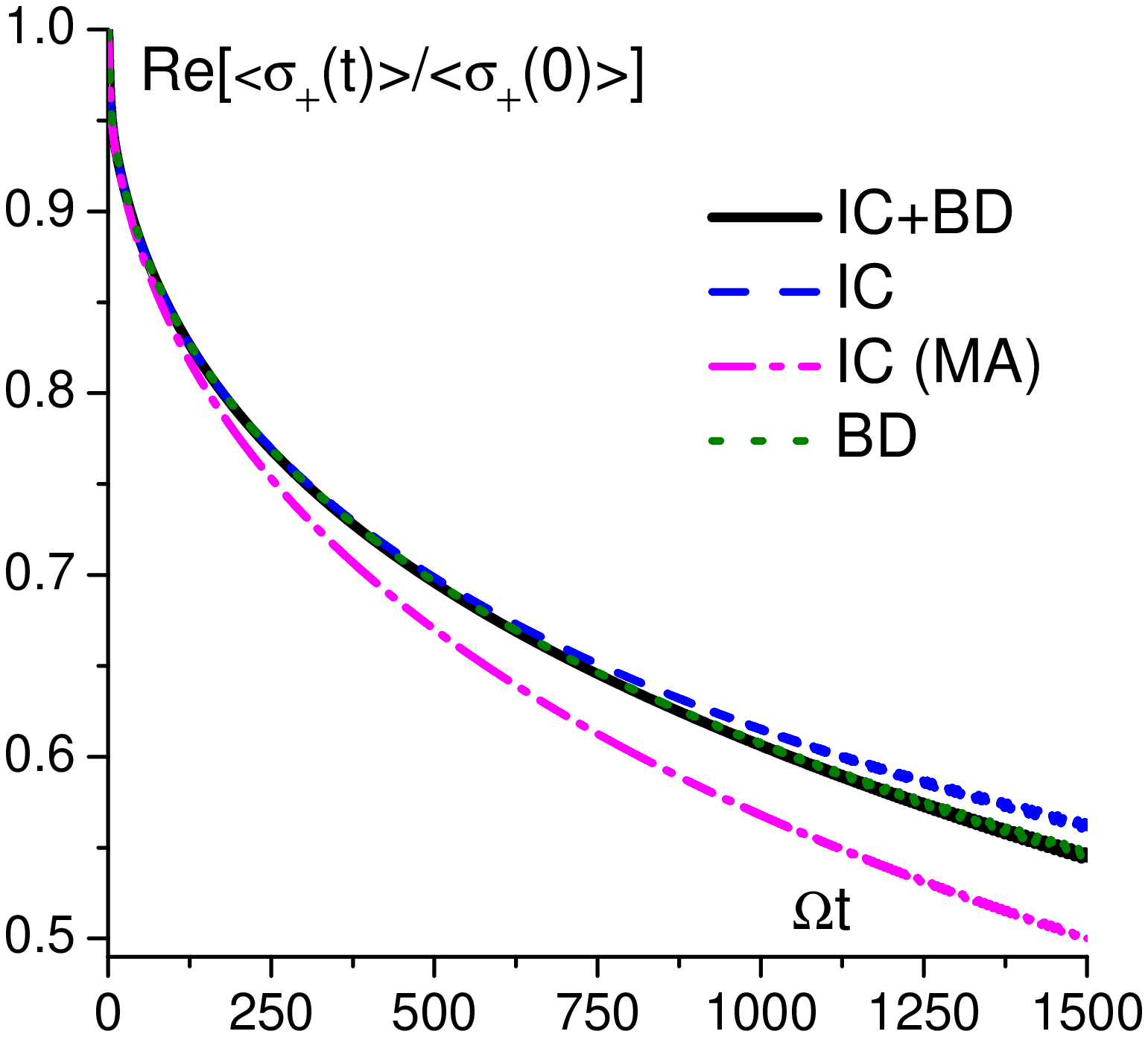}
		\hspace{1cm}
		\includegraphics[height=0.2\textheight,angle=0]{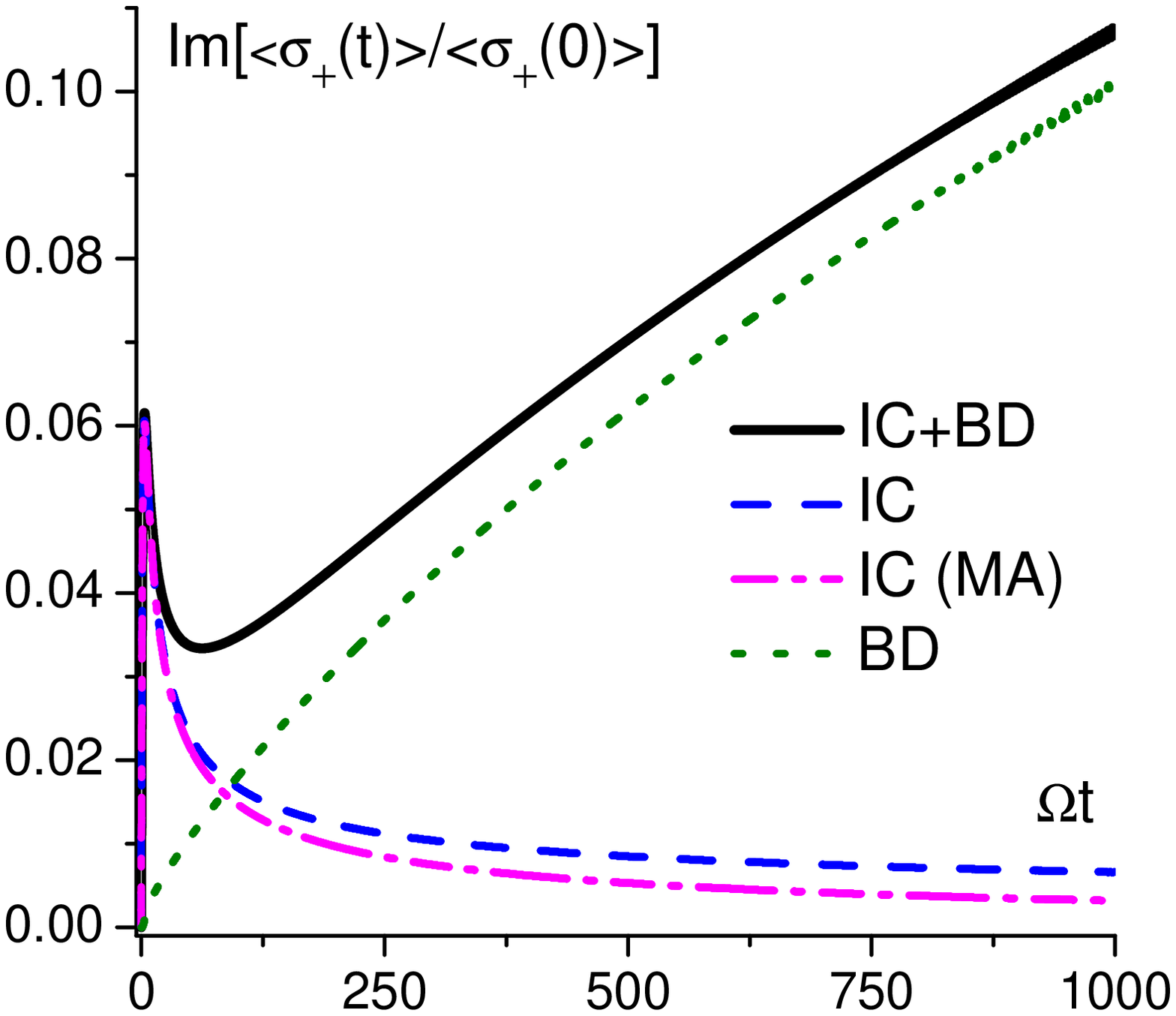}}
	\vspace{0.2cm}
	\centerline{\textbf{\footnotesize{HT}}\hspace{2mm}\includegraphics[height=0.2\textheight,angle=0]{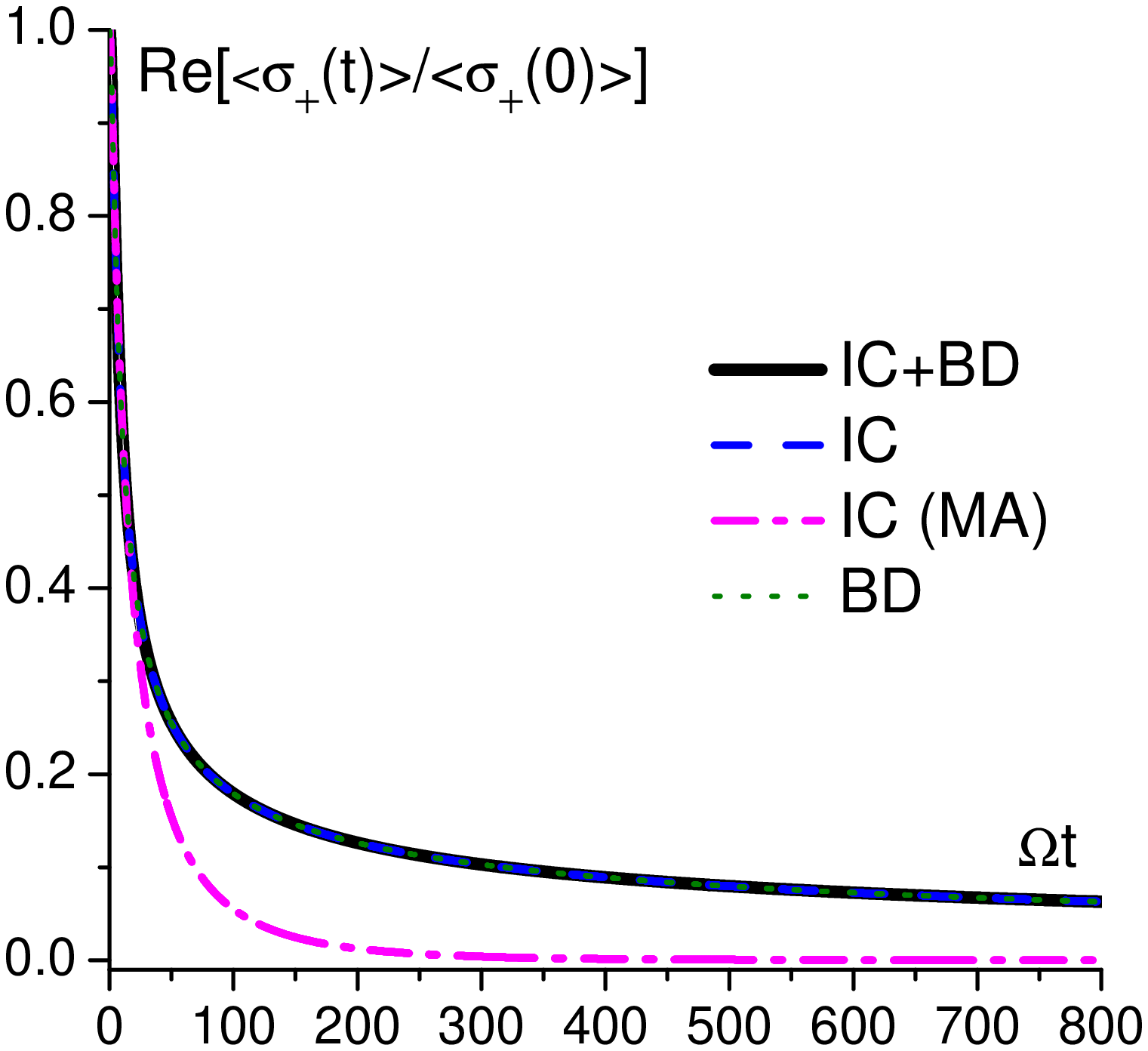}
		\hspace{1cm}
		\includegraphics[height=0.2\textheight,angle=0]{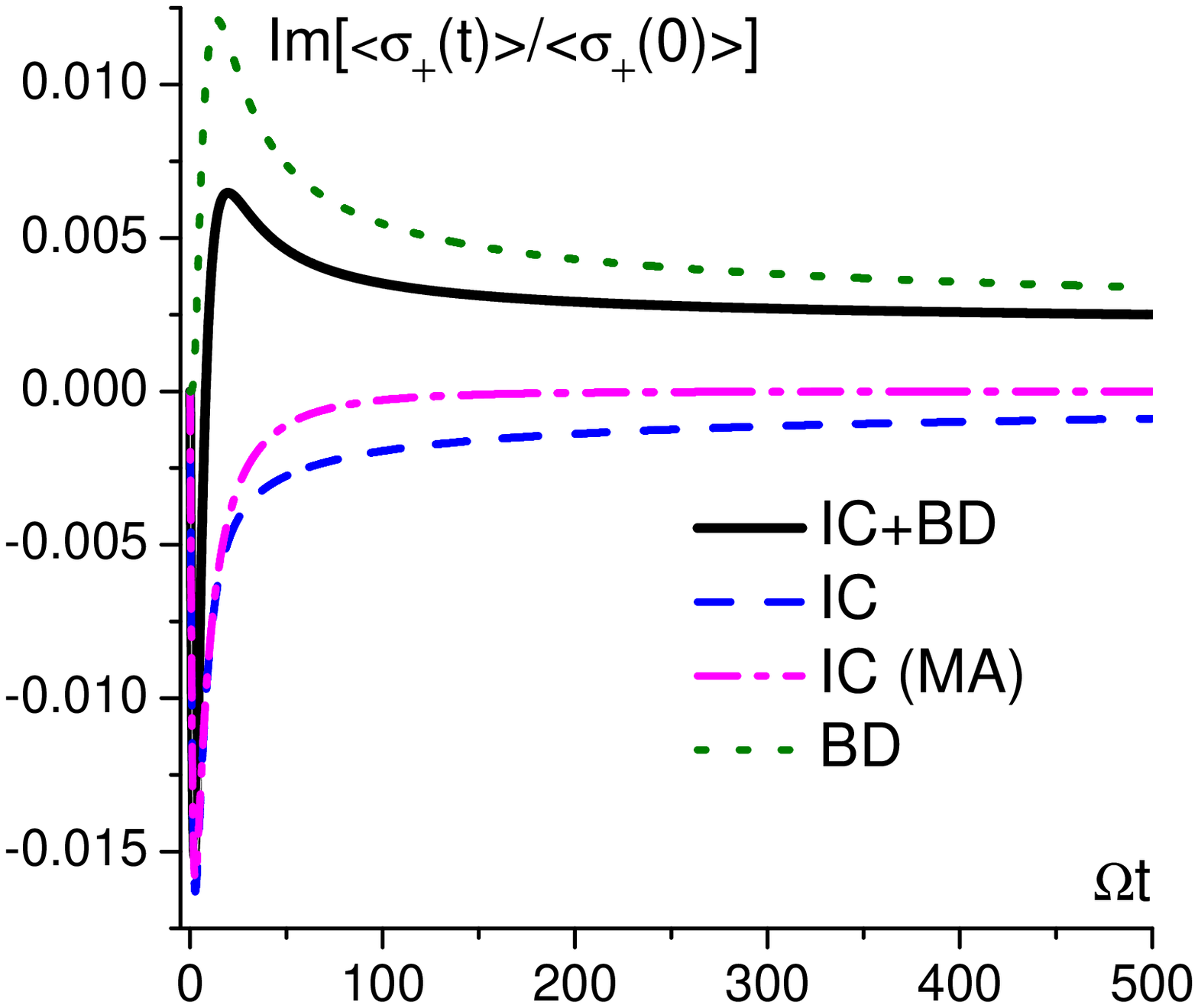}
}
	\caption{\small{The same as in Fig.~1 at the super-Ohmic coupling with $s=3/2$.}}
\end{figure*}

Now let us look at the left column of Fig.~2, where the time evolution of the real part of the generalized coherence is depicted at the super-Ohmic coupling.
One can see again a noticeable deviation of the non-Markovian dynamics from the exact results obtained within the MA for both low and high temperatures. Like for the Ohmic coupling in the high temperature limit (c.f. the lower panel of Fig.~1), there is a very small difference between the IC, BD and IC+BD cases. It implies that a non-Markovianity of the generic quantum kinetic equation itself dominates over any other factors of the qubit dynamics.

The situation changes drastically if one looks at the imaginary part of the system co\-he\-ren\-ce. Taking the BD into consideration leads to an uncontrollable increase of $\mbox{Im}[\langle\sigma_+(t)/\langle\sigma_+(0)\rangle]$ at low temperature. The picture does not improve  much even at high temperature, since the imaginary part of the system coherence tends to a saturation. It contradicts the conclusions made in Ref.~\cite{PRA2012} that the partial decoherence is admitted only at the strong super-Ohmic regime with $s>2$, while at smaller values of the ohmicity index the system coherence always tends to zero at large times. 

It is clear that such an unphysical behaviour of the system coherence needs a detailed explanation. 
Quantitatively, such explanation is provided in Section 7. Here we present only some ``intuitive'' reasons explaining why the non-Markovian quantum kinetic equation (though more accurate in a mathematical sense than the Markovian one) yields worse results as compared to the MA. 
In fact, a presence of the time convolution in Eq.~(\ref{kinetFin}) means that one has to account implicitly higher orders in the coupling constant when (formally) expanding the mean values of observables in series in $t'$. This is due to time dependence, which has to be evaluated with the evolution operator involving $V$. An alternative way \cite{BP-Book} is to pass from the time retarded master equation to the convolutionless one using the upward and backward evolution operators with a subsequent expansion in interaction term $V$. Such an approach is known to provide a much more controllable scheme for study of the qubit dynamics in case of the Janes-Cummings model (which is an exactly solvable one at zero temperature).  We believe that a main advantage of the non-Markovian quantum kinetic equations is dealt with description of the short-time system dynamics (especially in presence of alternating fields \cite{CMP2004,TMF2008}) rather than with the study of the system equilibration. 
%In some sence, one can build a ``bridge'' between the proposed approach and the quantum kinetic theory of many-particle system \cite{} provided that closer analogy can be stated only after a thorough investigation of the particular quantum open system.

\section{Reduced density matrix of the environment: exact result and approximate solutions}

\setcounter{equation}{0}

In this Section we estimate which accuracy the reduced density matrix $\widetilde{\varrho}_B(t)$ should be calculated with to ensure the reliable results for the BD. For the sake of simplicity, we assume the composite $(S+B)$ system to be uncorrelated at the initial time, 
$$
%\label{rho-comp-uncor}
\rho(0)= \rho_S\otimes \rho_B, \quad \rho_B=e^{-\beta
	H_B}/Z_B, \quad Z_B=\mbox{Tr}_B e^{-\beta
	H_B}.
$$
Using the expression (\ref{IP-rho-B-lin}) for the reduced density matrix of the environment in a linear approximation in interaction and Eq.~(\ref{tildeV1}) for $\tilde{V}(t)$, we obtain the following relation after application of the Kubo identity \cite{ZMR}:
\begin{eqnarray}\label{commut1}
\frac{1}{Z_B}\left[\tilde V_B(t'), e^{-\beta
	H_B}\right]=\frac{\beta}{Z_B}\int\limits_0^1 e^{-x\beta
	H_B}\left[H_B,\tilde V_B(t')\right]e^{x\beta H_B}\varrho_{B}\,dx.
\end{eqnarray}
The inner commutator in Eq.~(\ref{commut1}) can be presented as follows:
\begin{eqnarray}\label{commut2}
\left[H_B,\tilde
V_B(t')\right]=\langle\sigma_3\rangle\sum\limits_k\omega_k\left\{g_k
\exp(i\omega_k t')b^{\dagger}_k-g^*_k\exp(-i\omega_k t')b_k
\right\}.
\end{eqnarray}
Using the obvious relations 
\begin{eqnarray}\label{bx}\nonumber
&& b^{\dagger}_k(x)\equiv e^{-x\beta H_B}b^{\dagger}_k e^{-x\beta
	H_B}=e^{-\beta\omega_k x}b^{\dagger}_k,\\  
&& b_k(x)\equiv e^{-x\beta
	H_B}b_k e^{x\beta H_B}\,\,\,=\,\,\,e^{\beta\omega_k x}b_k,
\end{eqnarray}
and carrying out the integration over $x$, one can rewrite Eq.~(\ref{commut1}) in a more transparent form,
\begin{eqnarray}\label{commut3}
\frac{1}{Z_B}\left[\tilde V_B(t'), e^{\beta
	H_B}\right]=-\langle\sigma_3\rangle\sum\limits_k \left\{g_k
e^{i\omega_k t'}(e^{-\beta\omega_k}-1)b^{\dagger}_k+g^*_k
e^{-i\omega_k t'}(e^{\beta\omega_k}-1)b_k \right\}\rho_B.
\end{eqnarray}
The last step on our way to evaluate the reduced density matrix of the environment is the integration over time in Eq.~(\ref{IP-rho-B-lin}) that yields
\begin{eqnarray}\label{rho_B1}
\tilde\rho_B(t)\!=\!\left\{\! 1+
\langle\sigma_3\rangle\sum\limits_k \frac{1}{\omega_k}\left(g_k
(1-e^{i\omega_k t})(e^{-\beta\omega_k}-1)b^{\dagger}_k-g^*_k
(1-e^{-i\omega_k t})(e^{\beta\omega_k}-1)b_k
\right)\!\right\}\rho_B.
\end{eqnarray}
It is easy to verify using Eq.~(\ref{rho_B1}) that the reduced density matrix of the environment ensures correct dynamics for the bosonic mean values in the linear approximation in the coupling constant (c.f. Eqs.~(4)-(5) in Ref.~\cite{PRA2012}), namely:
\begin{eqnarray}\label{b(t)}\nonumber
%\langle \tilde b^{\dagger}_k\rangle^t=
&& \mbox{Tr}_B
b^{\dagger}_k\tilde\rho_B(t)=\mbox{Tr}_B
b^{\dagger}_k(t)\rho_B
=g_k\langle\sigma_3\rangle\frac{e^{i\omega_k t}-1}{\omega_k},\\
%\langle \tilde b_k\rangle^t=
&&\mbox{Tr}_B
b_k\tilde\rho_B(t)=\mbox{Tr}_B b_k(t)\rho_B=
g_k^*\langle\sigma_3\rangle\frac{e^{-i\omega_k t}-1}{\omega_k}.
\end{eqnarray}

However, it yields no correction to the nonequilibrium distribution function of the environment $n_k(t)=\mbox{Tr}_B\{ b^{\dagger}_k b_k
\tilde\rho_B(t)\}\equiv n_k^0$ (where $n_k^0=1/[\exp(\beta\omega_k)-1]$ denotes the equilibrium Bose distribution), as it can be easily checked using Eq.~(\ref{rho_B1}).
The above result is quite expected, since the reduced density matrix of the bath in the linear approximation in $g_k$ does not relax with time, as it has to follow from the general physical reasons. The observed
divergence of the generalized coherence at the super-Ohmic coupling (see Fig.~2) can be directly related to this unphysical behaviour of the approximated $\widetilde{\varrho}_B(t)$.

On the other hand, the quite simple form of the system Hamiltonian (\ref{H}) allows one to obtain not only the exact result for the coherence dynamics but also the exact expression for $\widetilde{\varrho}_B(t)$ \cite{CMP}. An inspection of Eq.~(2.17) in Ref.~\cite{CMP} shows that the relaxation of the bath density matrix is ensured by the non-equilibrium correction to the phonon energy
\bea\label{deltaEph}
\Delta\varepsilon_{ph}(t)=2\sum\limits_k\frac{|g_k|^2}{\omega_k}(1-\cos\omega_k t).
\eea
In the second order in the coupling constant it is easy to obtain an estimation for the non-equilibrium distribution function of phonons,
\begin{eqnarray}\label{nk(t)}
n_k(t)\sim\exp[-\beta\Delta\varepsilon_{ph}(t)]
n_k^0\approx\left(1-2\beta\sum\limits_q\frac{|g_q^2|}{\omega_q}(1-\cos\omega_q
t)\right)n_k^0,
\end{eqnarray}
which at least qualitatively is consistent \footnote{
	This is best seen in the high temperature limit, when the leading term of the Bose distribution function is $1/\beta\omega_k$. Thus the non-equilibrium correction to the phonon distribution function is proportional to $\sum\limits_q\frac{|g_q^2|}{\omega_k\omega_q}\left(1-\cos\omega_q t\right)$.} 
with the exact result \cite{PRA2012}
\begin{eqnarray}\label{nk_exact}
n_k(t)=n_k^0+\frac{2|g_k|^2}{\omega_k^2}(1-\cos\omega_k t).
\end{eqnarray}
To summarize, we would like to emphasize once more that it is not suffucient to calculate $\widetilde{\varrho}_B(t)$ in the linear approximation in interaction to describe the contribution of BD correctly. At least, the next order in the coupling constant is necessary. In such a case, one has to retain the corresponding high order terms in all the constituent parts of the generic quantum kinetic equation, which makes the problem almost unmanageable. 

On the other hand, the time decay of the reduced density matrix of the environment is determined by expression~ (\ref{deltaEph}). It follows from Ref.~\cite{PRA2012} that the time dependences of the vacuum and thermal contributions to the generalized decoherence are governed by similar expressions with higher  ohmicity indexes. It is straightforward to show that in the sub-Ohmic and Ohmic regimes, the typical decay time of $\varrho_B(t)$ is really smaller than the timescales at which the system coherence vanishes. Thus, in such cases the Markovian approximation can be well justified. In the strong super-Ohmic regime the situation changes significantly, and the above typical times become close to each other, making the non-Markovian description more reasonable. 

\section{Conclusions}

In this paper, we have performed an attempt to generalize the ordinary ZN projection scheme for derivation of the master equation of the open quantum system weakly coupled with its surroundings. This generalization consists in taking into account the intrinsic dynamics of the environment, which is supposed to be essential when one deals with very compact systems where surroundings of the quantum object can hardly be considered as a thermal bath \footnote{However we prefer to use the term ``bath dynamics'' throughout the paper when speaking about the time evolution of the reduced density matrix $\widetilde{\varrho}_B(t)$, since all the final averages are to be taken over the thermal bath.}.
We start from the chain of equations for the density matrices of the $S$- and $B$-subsystems, coupled to each other by the correlational part $\Delta\widetilde{\varrho}(t)$ of the total density matrix. The formal solution for $\Delta\widetilde{\varrho}(t)$ is of a very complicated structure involving all the orders in interaction and the projecting operators, acting in the Hilbert spaces of the open quantum system and the bath. Thus the generic master equation for $\widetilde{\varrho}_S(t)$ is also very complicated to be dealt with; besides, it is non-local in time.

We simplified an above mentioned equation, restricting ourselves by the second order in the coupling constant. 
However, a price paid for consideration of the bath dynamics is the extra term, which is found to be nonlinear in $\widetilde{\varrho}_S(t)$ and is vanishing in the Markovian limit. It allows one to make some allusions about a resemblance of the proposed approach to the concept of running (or dynamical) correlations, as it is usually takes place in the quantum kinetic theory \cite{ZMR,MorRop01,CMP2004,TMF2008}. If one neglects the dynamic equation for $\widetilde{\varrho}_B(t)$ (or, what is the same, adopts the decomposition (\ref{Zwanz-decomp-eq}) for the density matrix of the total system), he comes to the standard ZN projecting scheme \cite{BP-Book} without the term dealt with the BD.

To verify the elaborated scheme, we have applied this method to the open quantum system which is described by a very simple dephasing model. The non-Markovian quantum kinetic equation for the ge\-ne\-ra\-li\-zed coherence has been derived. In the Markovian approximation, the solution of this equation up to the 2-nd order in interaction coincides with the exact one obtained in \cite{PRA2012}. However, when the BD is taken into account, the numerical solution of the non-Markovian kinetic equation shows the unphysical behaviour: in the high-temperature limit at the super-Ohmic coupling with $s=3/2$ the imaginary part of ge\-ne\-ra\-li\-zed coherence $\langle\widetilde{\sigma}_+(t)\rangle$ diverges. The situation is not improved in the high temperature limit either, where the value of $\mbox{Im}\left[\langle\widetilde{\sigma}_+(t)\rangle\right]$ saturates at long times, that contradicts to the conception of total decoherence at this coupling regime \cite{PRA2012}.

We proposed the qualitative explanation of this fact, based on the exact and approximate forms of the reduced density matrix of the environment \cite{CMP}. It is shown that in the lowest approximation in interaction, the bath density matrix does not decay with time, and taking (at least) the next order into account is indispensable. 
However, it implies that the higher order terms  should be retained in the
generic master equation, making such a method of little use.

There is another approach to treat the intrinsic BD of the composite $(S+B)$ system. It is based upon taking into consideration the dynamical correlations in the system like it has been done in Refs.~\cite{MorRop01,ZMR}. These long-lived dynamical correlations, which are associated with the total energy conservation, play an important role in transition to the Markovian regime and subsequent equilibration of the system. In our recent article \cite{particles}, we have proposed a general scheme which can be applied to any open quantum system for investigation of the influence of running correlations. An application of this scheme to the above studied dephasing model will be the subject of our forthcoming paper.

\section*{Appendix A}
 \renewcommand{\theequation}{A.\arabic{equation}}
 \setcounter{equation}{0}
Let us look at the r.h.s. of Eq.~(\ref{IP-eq-S-weak}). Taking into account definition (\ref{Q-lin}) of the projector ${\mathcal{Q}^{(0)}(t)}$, we can express the integrand in a more detailed form:
	\beq
	\label{Term-Q}
	\begin{array}{c}
		\hspace*{-80pt}
		{\mathcal Q}^{(0)}(t'){\mathcal L}(t')
		\widetilde{\varrho}^{}_{S}(t')\varrho^{}_{B}(0)
		=
		{\mathcal L}(t')\widetilde{\varrho}^{}_{S}(t')\varrho^{}_{B}(0)
		\\[10pt]
		{}- \widetilde{\varrho}^{}_{S}(t')\,
		\text{Tr}^{}_{S}\left\{{\mathcal L}(t')
		\widetilde{\varrho}^{}_{S}(t')\varrho^{}_{B}(0)\right\}
		- \varrho^{}_{B}(0)\,
		\text{Tr}^{}_{B}\left\{
		{\mathcal L}(t')
		\widetilde{\varrho}^{}_{S}(t')\varrho^{}_{B}(0)
		\right\}.
	\end{array}
	\eeq
	Taking the trace over the variables of open quantum system, we can simplify a little the second term in (\ref{Term-Q}),
	\bea\nonumber
	\label{TrS}
	& &
	\text{Tr}^{}_{S}\left\{{\mathcal L}(t')
	\widetilde{\varrho}^{}_{S}(t')\varrho^{}_{B}(0)\right\}
	\\[5pt]
\nonumber	& &
	{}=
	\text{Tr}^{}_{S}\Big\{
	\underbrace{\left([\widetilde{V}(t'),
		\widetilde{\varrho}^{}_{S}(t')]\right)\varrho^{}_{B}(0)}_{\longrightarrow\ 0}
	+\widetilde{\varrho}^{}_{S}(t')
	\left([\widetilde{V}(t'),
	\varrho^{}_{B}(0)]\right)
	\Big\}
	\\[5pt]
	& &
	{}= {\mathcal L}^{}_{B}(t')\varrho^{}_{B}(0)
	\eea
	with operator ${\mathcal L}^{}_{B}(t')$ defined by Eq.~(\ref{H-int-B}).
	Analogously, it is possible to present a result of the trace action over the bath variables in the last term of Eq.~(\ref{Term-Q}) as follows:
	\bea\nonumber	\label{TrB}
	& &
	\text{Tr}^{}_{B}\left\{{\mathcal L}(t')
	\widetilde{\varrho}^{}_{S}(t')\varrho^{}_{B}(0)\right\}
	\\[5pt]
\nonumber	& &
	{}=
	\text{Tr}^{}_{B}\Big\{
	\left([\widetilde{V}(t'),
	\widetilde{\varrho}^{}_{S}(t')]\right)\varrho^{}_{B}(0)
	+\underbrace{\widetilde{\varrho}^{}_{S}(t')
		\left([\widetilde{V}(t'),
		\varrho^{}_{B}(0)]\right)}_{\longrightarrow\ 0}
	\Big\}
	\\[5pt]
	& &
	{}= {\mathcal L}^{(0)}_{S}(t')\widetilde{\varrho}^{}_{S}(t')
	\eea
	where we have introduced the denotations
	\beq
	\label{L-0-S}
	{\mathcal L}^{(0)}_{S}(t)A=\left[\widetilde{V}^{(0)}_{S}(t),A\right],
	\qquad
	\widetilde{V}^{(0)}_{S}= \text{Tr}^{}_{B}
	\left\{
	\widetilde{V}(t)\varrho^{}_{B}(0)
	\right\}.
	\eeq
	
	Collecting the results obtained in (\ref{TrS})-(\ref{TrB}), we can rewrite Eq.~(\ref{Term-Q}) as follows:
	\beq
	\label{Term-Q-1}
	{\mathcal Q}^{}_{0}(t'){\mathcal L}(t')
	\widetilde{\varrho}^{}_{S}(t')\varrho^{}_{B}(0)
	={\mathcal L}(t')\widetilde{\varrho}^{}_{S}(t')\varrho^{}_{B}(0)
	-\widetilde{\varrho}^{}_{S}(t'){\mathcal L}^{}_{B}(t')\varrho^{}_{B}(0)
	-\varrho^{}_{B}(0){\mathcal L}^{(0)}_{S}(t')\widetilde{\varrho}^{}_{S}(t').
	\eeq
	Now we have all the components to calculate the integrand in Eq.~(\ref{IP-eq-S-weak}): 
	\bea\label{Int}\nonumber
	& &
	\hspace*{-50pt}
	\mbox{Integrand}\equiv
	\text{Tr}^{}_{B}
	\left\{
	{\mathcal L}(t){\mathcal Q}^{}_{0}(t'){\mathcal L}(t')
	\widetilde{\varrho}^{}_{S}(t')\varrho^{}_{B}(0)
	\right\}
	\\[10pt]
	& &
\nonumber	{}=\underbrace{\text{Tr}^{}_{B}
		\left\{
		{\mathcal L}(t)
		{\mathcal L}(t')\widetilde{\varrho}^{}_{S}(t')\varrho^{}_{B}(0)
		\right\}}_{I}
	\\[10pt]
\nonumber	& &
	{}-\underbrace{\text{Tr}^{}_{B}
		\left\{
		{\mathcal L}(t)
		\widetilde{\varrho}^{}_{S}(t'){\mathcal L}^{}_{B}(t')\varrho^{}_{B}(0)
		\right\}}_{II}
	\\[10pt]
	& &
	{}-\underbrace{\text{Tr}^{}_{B}
		\left\{
		{\mathcal L}(t)
		\varrho^{}_{B}(0){\mathcal L}^{(0)}_{S}(t')\widetilde{\varrho}^{}_{S}(t')
		\right\}}_{III}.
	\eea
	Let us transform the terms $II$ and $III$. It is evident that  one can split $II$ into a sum of two terms, one of
	which vanishes. We thus have
\bea\label{IntII}
	II= \text{Tr}^{}_{B}\left\{
	\left({\mathcal L}(t)\widetilde{\varrho}_{S}(t')\right)
	{\mathcal L}^{}_{B}(t')\varrho^{}_{B}(0)
	\right\}=
	\text{Tr}^{}_{B}\left\{
	\left(\widetilde{V}(t)\widetilde{\varrho}_{S}(t')
	-\widetilde{\varrho}_{S}(t')\widetilde{V}(t)
	\right)
	{\mathcal L}^{}_{B}(t')\varrho^{}_{B}(0)
	\right\}.
	\eea
	We now introduce an operator acting in the Hilbert space of $S$:
	\beq
	\label{R}
	R^{}_{S}(t,t')=\text{Tr}^{}_{B}\left\{
	\widetilde{V}(t){\mathcal L}^{}_{B}(t')\varrho^{}_{B}(0)
	\right\}=\text{Tr}^{}_{B}\left\{
	\left[ \widetilde{V}(t),
	\widetilde{V}^{}_{B}(t')\right]\varrho^{}_{B}(0)
	\right\}.
	\eeq
Using this definition, it is possible to express the term $II$ as follows:
	\beq
	\label{II}
	II= \left[ R^{}_{S}(t,t'),\widetilde{\varrho}_{S}(t')\right].
	\eeq
	The term $III$ can also be splitted into two summands, one of which vanishes.
	As a result we have
	\beq
	\label{III}
	III= {\mathcal L}^{(0)}_{S}(t){\mathcal L}^{(0)}_{S}(t')
	\widetilde{\varrho}_{S}(t').
	\eeq

Combining the above results, Eq.~(\ref{IP-eq-S-weak}) takes the form
\bea
\label{master-1}
& &
\hspace*{-40pt}
\frac{\partial \widetilde{\varrho}^{}_{S}(t)}{\partial t}=
-i{\mathcal L}^{}_{S}(t)\widetilde{\varrho}^{}_{S}(t)
- \int^{t}_{0}dt'\, \text{Tr}^{}_{B}
\left\{
{\mathcal L}(t)
{\mathcal L}(t')\widetilde{\varrho}^{}_{S}(t')\varrho^{}_{B}(0)
\right\}
\nonumber\\[10pt]
& &
{}+
\int^{t}_{0}dt'\,
\left\{
\left[ R^{}_{S}(t,t'),\widetilde{\varrho}_{S}(t')\right]
+{\mathcal L}^{(0)}_{S}(t){\mathcal L}^{(0)}_{S}(t')
\widetilde{\varrho}_{S}(t')
\right\}.
\eea

Let us now consider the first term on the r.h.s. of	Eq.~(\ref{master-1}). Taking into account (\ref{H-int-S}) and  definition (\ref{L-def}) of the Liouville operator ${\mathcal L}_S(t)$, we obtain
	\bea\label{tildeVS1}
	\widetilde{V}^{}_{S}(t)= \text{Tr}^{}_{B}
	\left\{\widetilde{V}(t)\varrho^{}_{B}(0)\right\}
	-i\int^{t}_{0} dt'\, \text{Tr}^{}_{B}
	\left\{\widetilde{V}(t){\mathcal L}^{}_{B}(t')\varrho^{}_{B}(0)
	\right\}.
	\eea
	Recalling Eqs.~(\ref{L-0-S}) and (\ref{R}), it gives
	\beq
	\label{V-S-lin}
	\widetilde{V}^{}_{S}(t)=\widetilde{V}^{(0)}_{S}(t)
	-i\int^{t}_{0}dt'\,
	R^{}_{S}(t,t'),
	\eeq
or, collecting all the above terms, it yeilds the expression
	\beq
	\label{Drift}
		-i{\mathcal L}^{}_{S}(t)\widetilde{\varrho}^{}_{S}(t)=
		-i{\mathcal L}^{(0)}_{S}(t)\widetilde{\varrho}^{}_{S}(t)
		- \int^{t}_{0}dt'\,\left[R^{}_{S}(t,t'),\widetilde{\varrho}^{}_{S}(t)\right],
	\eeq
which finalizes the form of Eq.~(\ref{master-fin}).

\section*{Appendix B}
\renewcommand{\theequation}{B.\arabic{equation}}
\setcounter{equation}{0}

To derive the master equation within the ZN scheme, we may again start from Eq.~(\ref{IP-eq-S-1}).
This equation does not depend on a particular decomposition of the
total statistical operator $\widetilde{\varrho}(t)$. Substituting (\ref{Zw-IP-decomp}) in the generic Eq.~(\ref{IP-eq-S-1}), we obtain
\beq
\label{IP-eq-S-1b}
\frac{\partial\widetilde{\varrho}^{}_{S}(t)}{\partial t}=
- i{\mathcal L}^{(0)}_{S}(t)\widetilde{\varrho}^{}_{S}(t)
-i\,\text{Tr}^{}_{B}\left[\widetilde{V}(t),\delta\widetilde{\varrho}(t)\right],
\eeq
where ${\mathcal L}^{(0)}_{S}(t)$ is defined by (\ref{L-0-S}).

Let us derive the equation of motion for $\delta\widetilde{\varrho}(t)$
using Eq.~(\ref{Zw-IP-decomp}). We perform it in a similar way as it has been done in the Section 3.
Taking the time derivative of Eq.~(\ref{Zw-IP-decomp}), one gets
$$
\frac{\partial}{\partial t}\,\delta\widetilde{\varrho}=
\frac{\partial \widetilde{\varrho}}{\partial t}
- \varrho^{}_{B}(0)\,\frac{\partial \widetilde{\varrho}^{}_{S}}{\partial t}=-i\left[\widetilde{V}(t),\widetilde{\varrho}(t)\right]
+ i\varrho^{}_{B}(0)\,\text{Tr}^{}_{B}
\left[\widetilde{V}(t),\widetilde{\varrho}(t)\right].
$$
Again using decomposition (\ref{Zw-IP-decomp}), we obtain
\bea
\label{Eq-delta}
& &
\frac{\partial}{\partial t}\,\delta\widetilde{\varrho}(t)=
-i\left[\widetilde{V}(t),\widetilde{\varrho}^{}_{S}(t)\varrho^{}_{B}(0)\right]
-i\left[\widetilde{V}(t),\delta\widetilde{\varrho}(t)\right]
\nonumber\\[10pt]
& &
{}+i
\varrho^{}_{B}(0)\,\text{Tr}^{}_{B}
\left[\widetilde{V}(t),\widetilde{\varrho}^{}_{S}(t)\varrho^{}_{B}(0)\right]
+i \varrho^{}_{B}(0)\,\text{Tr}^{}_{B}
\left[\widetilde{V}(t),\delta\widetilde{\varrho}(t)\right].
\eea
Let us now introduce the Zwanzig projecting operators:
\beq
\label{P-Zw}
{\mathcal P}^{}_{Z}A= \varrho^{}_{B}(0)\,\text{Tr}^{}_{B}A,
\qquad
{\mathcal Q}^{}_{Z}=1 - {\mathcal P}^{}_{Z}.
\eeq
If one compares the form of Zwanzig projecting operator (\ref{P-Zw}) with that appearing in the generalized scheme (\ref{P-Q}), he will note that ignoring the time evolution of the reduced density matrix of the environment makes the term $\widetilde{\varrho}_S(t)\mbox{Tr}_S A$ (dealt with $\partial\widetilde{\varrho}_B(t)/\partial t$) to vanish.
Using the definition (\ref{P-Zw}) of the Zwanzig projecting operators and the obvious equality 
$
{\mathcal P}^{}_{Z}\,\delta\widetilde{\varrho}(t)=0
$,
Eq.~(\ref{Eq-delta}) can be rewritten in a more compact form,
\beq
\label{Eq-delta-Z}
\left(
\frac{\partial}{\partial t}
+ i {\mathcal Q}^{}_{Z}\,{\mathcal L}(t)\,{\mathcal Q}^{}_{Z}
\right)\delta\widetilde{\varrho}(t)=
-i{\mathcal Q}^{}_{Z}\,{\mathcal L}(t)\,
\widetilde{\varrho}^{}_{S}(t)\varrho^{}_{B}(0).
\eeq
A formal solution of (\ref{Eq-delta-Z}) is given by (cf. Eq.~(\ref{Del-rho-formal}))
\beq
\label{delta-rho-Z}
\delta\widetilde{\varrho}(t)=
-i\int^{t}_{0} dt'\,
{\mathcal U}^{}_{Z}(t,t'){\mathcal Q}^{}_{Z}\,{\mathcal L}(t')\,
\widetilde{\varrho}^{}_{S}(t')\varrho^{}_{B}(0),
\eeq
where
\beq
\label{Zw-U-I}
{\mathcal U}^{}_{Z}(t,t')=\exp^{}_{+}
\left\{
-i\int^{t}_{t'}d\tau\,
{\mathcal Q}^{}_{Z}{\mathcal L}(\tau){\mathcal Q}^{}_{Z}
\right\}.
\eeq
Substituting (\ref{delta-rho-Z}) into (\ref{IP-eq-S-1b}), we arrive
at the ZN master equation (\ref{Zw-master-I}).

In the weak coupling approximation, we one can set ${\mathcal U}^{}_{Z}(t,t')=1$ like it has been done in Section 3. Noting that
\bea\label{QZ-action}\nonumber
& &
{\mathcal Q}^{}_{Z}\,{\mathcal L}(t')\,
\widetilde{\varrho}^{}_{S}(t')\varrho^{}_{B}(0)=
{\mathcal L}(t')\,
\widetilde{\varrho}^{}_{S}(t')\varrho^{}_{B}(0)
- \varrho^{}_{B}(0)\,
\text{Tr}^{}_{B}\left\{
{\mathcal L}(t')\,
\widetilde{\varrho}^{}_{S}(t')\varrho^{}_{B}(0)
\right\}
\\[10pt]
& &
{}= {\mathcal L}(t')\,
\widetilde{\varrho}^{}_{S}(t')\varrho^{}_{B}(0)
- \varrho^{}_{B}(0)\,{\mathcal L}^{(0)}_{S}(t')
\widetilde{\varrho}^{}_{S}(t'),
\eea
we thus have
\bea\label{B8}
\text{Tr}^{}_{B}
\left\{
{\mathcal L}(t){\mathcal Q}^{}_{Z}\,{\mathcal L}(t')\,
\widetilde{\varrho}^{}_{S}(t')\varrho^{}_{B}(0)
\right\}=
\text{Tr}^{}_{B}
\left\{
{\mathcal L}(t){\mathcal L}(t')\,
\widetilde{\varrho}^{}_{S}(t')\varrho^{}_{B}(0)
\right\}
- {\mathcal L}^{(0)}_{S}(t){\mathcal L}^{(0)}_{S}(t')
\widetilde{\varrho}^{}_{S}(t').
\eea
Substituting the above result into Eq.~(\ref{Zw-master-I}), we obtain the master equation (\ref{ZW-master-wI}) in the ZN scheme, which is valid up to the second order in interaction.

\section*{Appendix C}
\renewcommand{\theequation}{C.\arabic{equation}}
\setcounter{equation}{0}

Using expression (\ref{tildeV1}) for the interaction operator
$\tilde V(t)$ and taking into account the permutation relations
between the matrices $\sigma_3$ and $\sigma_+$, one can easily
obtain the first term in Eq.~(\ref{kinet1}),
\bea\label{one}(1)=2i\left\langle\widetilde\sigma_+\right\rangle^t_{S}\sum\limits_k\langle{\cal F}_k(t)\rangle_B.\eea Eq.~(\ref{one})
implies calculation of the ``anomalous'' thermal bath
averages $\langle b^{\dagger}_k\rangle_B$ and $\langle
b_k\rangle_B$, which are non-zero if the initial value of the reduced density matrix of the environment differs from equilibrium, $\rho_B(0)\ne\exp(-\beta H_B)/Z_B$. The above mentioned bath averages of these phonon operators as well as their products are calculated in Appendix D.

To obtain the integrands (2) and (2') in the kinetic equation (\ref{kinet1}), which describe the contribution of the intrinsic dynamics of the environment, let us calculate the commutator with interaction potentials, entering in expression (\ref{R}). This commutator can be presented as follows:
\bea\label{commutVVB}\left[\tilde V(t),\tilde
V_B(t')\right]=\sigma_3\langle\sigma_3\rangle\sum\limits_{k,k'}\left[{\cal
	F}_k(t),{\cal F}_{k'}(t')\right],\eea
where
\bea\label{commutFkFk'}\sum\limits_{k,k'}\left[{\cal F}_k(t),{\cal
	F}_{k'}(t')\right]=\sum\limits_{k,k'}\left\{g_k^*(t)g_{k'}(t')\left[b_k,b^{\dagger}_{k'}\right]+g_k(t)g^*_{k'}(t')\left[b_k^{\dagger},b_{k'}\right]\right\}\\
=2 i \mbox{Im}\sum\limits_k|g_k|^2 \exp(-i\omega_k(t-t'))=-2
i\sum\limits_k|g_k|^2\sin[\omega_k(t-t')].\eea

Passing from the discrete bath modes to the continuum limit and introducing the spectral density $J(\omega)$ in the usual fashion,
\bea\label{J(w)}
\sum\limits_k 4|g_k|^2 f(\omega_k)=\int\limits_0^{\infty}d\omega J(\omega) f(\omega),
\eea
one can express the integrands (2) and (2') in the following way:
\bea\label{I2}\nonumber
(2)=4
i\langle\sigma_3\rangle\left\{\langle\widetilde\sigma_+\rangle^{t}_{S}-\langle\widetilde\sigma_+\rangle^{t'}_{S}
\right\}\sum\limits_k
|g_k|^2\sin[\omega_k(t-t')]\\=i\langle\sigma_3\rangle\left\{\langle\widetilde\sigma_+\rangle^{t}_{S}-\langle\widetilde\sigma_+\rangle^{t'}_{S}\right\}\int\limits_0^{\infty}J(\omega)\sin[\omega(t-t')]d\omega.\eea
It is seen that expression~(\ref{I2}) contains an
imaginary unit, and is expected (along with the ``quasi-free''
term (\ref{one})) to contribute to the shift of the qubit frequency
\cite{PRA2012}.

To obtain the expression for integrand (3), at the beginning let us calculate the inner commutator in Eq.~(\ref{kinet1}),
\bea\label{I3K1} \left[\tilde
V(t'),\sigma_+\right]=2\sigma_+\sum\limits_{k'}{\cal F}_{k'}(t').
\eea
Thus, the total commutator in (\ref{kinet1}) can be presented
as follows: \bea\label{I3K2}\nonumber \left[\tilde
V(t),\left[\tilde V(t'),\sigma_+
\right]\right]=2\sum\limits_{k,k'}\left\{\sigma_3\sigma_+{\cal
	F}_{k}(t){\cal F}_{k'}(t')-\sigma_+\sigma_3{\cal F}_{k'}(t'){\cal
	F}_{k}(t)\right\}\\
=2\sigma_+\sum\limits_{k,k'}\left[{\cal F}_{k}(t),{\cal
	F}_{k'}(t')\right]_+. \eea 

The anticommutator in Eq.~(\ref{I3K2})
has the following form: \bea\label{anti}\nonumber \left[{\cal
	F}_{k}(t),{\cal
	F}_{k'}(t')\right]_+=\sum\limits_k\left\{g_k(t)g_k^*(t')+g_k^*(t)g_k(t')
\right\}\\
\nonumber+2 \sum\limits_{k,k'}\left\{g_k(t)g_{k'}(t')b^{\dagger}_k
b^{\dagger}_{k'}+g_k^*(t)g_{k'}^*(t')b_k b_{k'}\right\}\\+2
\sum\limits_{k,k'}\left\{g_k(t)g^*_{k'}(t')+g_k(t')g^*_{k'}(t)
\right\}b^{\dagger}_k b_{k'}.\eea Having multiplied
Eq.~(\ref{anti}) by $2\sigma_+$ and evaluated the corresponding
thermal bath mean values (see Appendix D for details), one can obtain the expression for
integrand (3).

The last integrand in Eq.~(\ref{kinet1}) can be easily calculated in a similar way using the expression
(\ref{one}) for the inner commutator. The final results looks as
\bea\label{I4}
(4)=4\left\langle\sigma^+\right\rangle^{t'}_{\widetilde S}
\sum\limits_{k,k'}\left\langle {\cal F}_k(t)\right\rangle_B
\left\langle{\cal F}_{k'}(t')_B\right\rangle.\eea
The thermal bath mean values $\left\langle{\cal F}_{k}(t)\right\rangle_B$ will be
calculated in Appendix D.

\section*{Appendix D}
\renewcommand{\theequation}{D.\arabic{equation}}
\setcounter{equation}{0}

Let us suppose that at the initial time $t=0$ the statistical operator
of the total system ``qubit+bath'' can be presented as a direct
product \bea\label{rhoNull}
\tilde\rho(0)=|\psi\rangle|\langle\psi|\bigotimes\rho_B(\psi).
\eea It follows from Eq.~(\ref{rhoNull}) that initially the qubit
was prepared in the pure state with the state-vector $|\psi\rangle=a_0|0\rangle+a_1|1\rangle$. Taking use of Eqs.~(20)--(21) of
Ref.~\cite{PRA2012}, one can present the bath operator $\rho_B(0)$ in
the following way: \bea\label{rhoBNull}\nonumber&&
\rho_B(0)\equiv\rho_B(\psi)=\rho_B^{(-)}+\rho_B^{(+)},\\\nonumber
&&\rho_B^{(+)}=\frac{|a_1|^2\exp(-\beta\omega_0/2)\exp(-\beta
	H_B^{(+)})}{|a_0|^2\exp(\beta\omega_0/2)Z_B^{(-)}+|a_1|^2\exp(-\beta\omega_0/2)Z_B^{(+)}}\\
&&\rho_B^{(-)}=\frac{|a_0|^2\exp(\beta\omega_0/2)\exp(-\beta
	H_B^{(-)})}{|a_0|^2\exp(\beta\omega_0/2)Z_B^{(-)}+|a_1|^2\exp(-\beta\omega_0/2)Z_B^{(+)}},
\eea
where \bea\label{HBpm} H_B^{(\pm)}=\sum\limits_k\omega_k
b^{\dagger}_k
b_k\pm\sum\limits_k(g_k b^{\dagger}_k+g^*_k b_k),\\
Z_B^{(\pm)}=\mbox{Tr}_B\exp(-\beta H_B^{(\pm)}). \eea \noindent
Calculation of the thermal bath averages can be simplified
considerably if one applies a unitary transformation of bosonic
operators with \bea\label{Upm} U_{\pm}=\exp\left\{
\pm\sum\limits_k\left(\frac
{g_k}{\omega_k}b^{\dagger}_k-\frac{g^*_k}{\omega_k}b_k \right)
\right\}. \eea In particular, it can be easily verified that

\bea\label{unitHpm} U_{\pm} H^{(\pm)}_B
U^{-1}_{\pm}=H_B-\sum\limits_k\frac{|g_k|^2}{\omega_k}.\eea
Consequently, \bea\label{ZB}\nonumber &&
Z_B^{(\pm)}=\mbox{Tr}_B\exp\left(-\beta
H^{(\pm)}_B\right)=\mbox{Tr}_B U^{-1}_{\pm}U_{\pm} \exp(-\beta
H^{(\pm)}_B)\\
&&=\mbox{Tr}_B U_{\pm} \exp(-\beta
H^{(\pm)}_B)U^{-1}_{\pm}=\exp\left(\beta\sum\limits_k\frac{|g_k|^2}{\omega_k}
\right)\mbox{Tr}_B\exp\left(-\beta H_B\right).\eea 
The $c$-factor $\exp(\beta\sum_k\frac{|g_k|^2}{\omega_k})$ is being
cancelled both in the numerator and the denominator of
Eq.~(\ref{rhoBNull}) yielding \bea\label{rhoBpm}\nonumber
&&\rho_B^{(+)}=\frac{|a_1|^2\exp(-\beta\omega_0/2)}{|a_0|^2\exp(\beta\omega_0/2)+|a_1|^2\exp(-\beta\omega_0/2)}
\frac{\exp(-\beta H_B)}{Z_B},
\\
&&\rho_B^{(-)}=\frac{|a_0|^2\exp(\beta\omega_0/2)}{|a_0|^2\exp(\beta\omega_0/2)+|a_1|^2\exp(-\beta\omega_0/2)}
\frac{\exp(-\beta H_B)}{Z_B}. \eea On the other hand,
transformation rules the for creation/annihilation ope\-ra\-tors
look as follows \cite{TMF}: \bea\label{unitb}\nonumber && U_+ b_k
U^{-1}_+\equiv\tilde b_k=b_k-\frac{g_k}{\omega_k},\quad  U_+
b_k^{\dagger} U^{-1}_+\equiv\tilde
b_k^{\dagger}=b_k^{\dagger}-\frac{g_k^*}{\omega_k},\\
&& U_- b_k U^{-1}_-\equiv\bar b_k=b_k+\frac{g_k}{\omega_k},\quad
U_- b_k^{\dagger} U^{-1}_-\equiv\bar
b_k^{\dagger}=b_k^{\dagger}+\frac{g_k^*}{\omega_k}.\eea It it seen
from Eqs.~(\ref{rhoBpm})-(\ref{unitb}) that evaluation of the mean
values like $\mbox{Tr}_B\rho^{(\pm)}_B b_k$,
$\mbox{Tr}_B\rho^{(\pm)}_B b_k^{\dagger}$ implies calculation of
averages $\mbox{Tr}_B\rho_B \tilde b_k$, $\mbox{Tr}_B\rho_B \tilde
b_k^{\dagger}$ (or, corresponding\-ly, $\mbox{Tr}_B\rho_B \bar
b_k$, $\mbox{Tr}_B\rho_B \bar b_k^{\dagger}$). The result looks
quite simple, \bea\label{meanb_B}\nonumber && \mbox{Tr}_B
\{\rho_B\bar b_k\}=\frac{g_k}{\omega_k},\quad
\mbox{Tr}_B\{\rho_B\bar
b_k^{\dagger}\}=\frac{g_k^*}{\omega_k},\\
&& \mbox{Tr}_B \{\rho_B\tilde b_k\}=-\frac{g_k}{\omega_k},\quad
\mbox{Tr}_B\{\rho_B\tilde
b_k^{\dagger}\}=-\frac{g_k^*}{\omega_k}.\eea

Thus, taking into account Eq.~(\ref{meanb_B}) along with
(\ref{rhoBpm}), it is possible to write down the following result
for thermal bath mean values: \bea\label{meanb}\nonumber
\mbox{Tr}_B \rho_B(0)
b_k=\underbrace{\frac{|a_0|^2\exp(\beta\omega_0/2)-|a_1|^2\exp(-\beta\omega_0/2)}{|a_0|^2\exp(\beta\omega_0/2)+|a_1|^2\exp(-\beta\omega_0/2)}}_{A(\psi)}
\frac{g_k}{\omega_k}=A(\psi)\frac{g_k}{\omega_k},\\ \mbox{Tr}_B
\rho_B(0)
b_k^{\dagger}=\frac{|a_0|^2\exp(\beta\omega_0/2)-|a_1|^2\exp(-\beta\omega_0/2)}{|a_0|^2\exp(\beta\omega_0/2)+|a_1|^2\exp(-\beta\omega_0/2)}
\frac{g_k^*}{\omega_k}=A(\psi)\frac{g_k^*}{\omega_k}.\eea The
result (\ref{meanb}) can be presented in a slightly different form.
Taking into account the normalization condition $|a_0|^2+|a_1|^2=1$
as well as the definition for the mean inversion population
$\langle\sigma_3\rangle=|a_0|^2-|a_1|^2$, it is possible to
express the factor $A(\psi)$ in Eqs.~(\ref{meanb}) via
$\langle\sigma_3\rangle$ as follows:\bea\label{A} A(\psi)=
\frac{\exp(\beta\omega_0/2)(1+\langle\sigma_3\rangle)-\exp(-\beta\omega_0/2)(1-\langle\sigma_3\rangle)}{\exp(\beta\omega_0/2)(1+\langle\sigma_3\rangle)+
	\exp(-\beta\omega_0/2)(1-\langle\sigma_3\rangle)}.\eea Expression
(\ref{A}) converts to $A(\psi)=\tanh(\beta\omega_0/2)$ at equal
initial populations of the ground and the excited states. On the
other hand, it equals to unity at the complete inversion
population, when $\langle\sigma_3\rangle=-1$.

In a similar way, it is straightforward to obtain the thermal bath
averages for other combinations of the bosonic operators, namely:
\bea\label{meanbb}\nonumber \mbox{Tr}_B\{ \rho_B \bar b_k\bar
b_{k'}\}=\mbox{Tr}_B\{ \rho_B \tilde b_k\tilde b_{k'}\}=\frac{g_k
	g_{k'}}{\omega_k\omega_{k'}},\\
\nonumber \mbox{Tr}_B\{ \rho_B \bar b_k^{\dagger}\bar
b_{k'}^{\dagger}\}=\mbox{Tr}_B\{ \rho_B \tilde b_k^{\dagger}\tilde
b_{k'}^{\dagger}\}=\frac{g_k^*
	g_{k'}^*}{\omega_k\omega_{k'}},\\
\mbox{Tr}_B\{ \rho_B \bar b_k^{\dagger}\bar b_{k'}\}=\mbox{Tr}_B\{
\rho_B \tilde b_k^{\dagger}\tilde
b_{k'}\}=n_k^0\delta_{kk'}+\frac{g_k^*
	g_{k'}}{\omega_k\omega_{k'}}.\eea 
Here $n_k^0$ means the equilibrium Bose distribution function
and $\delta_{kk'}$ denotes the
Kronecker $\delta$-symbol. It also follows from
Eq.~(\ref{rhoBNull}) that expressions (\ref{meanbb}) are nothing
but the mean values $\mbox{Tr}_B\{\rho_B(0) b_k b_{k'}\}$,
$\mbox{Tr}_B\{\rho_B(0) b_k^{\dagger} b_{k'}^{\dagger}\}$ and
$\mbox{Tr}_B\{\rho_B(0) b_k^{\dagger} b_{k'}\}$ since
corresponding ``barred'' and ``tilded'' averages are equal to each
other.

\end{document}